# Discovery of novel antimicrobial peptides with notable antibacterial potency by a LLM-based foundation model

Jike Wang[1,2,#], Jianwen Feng[4,#], Yu Kang[1,#], Peichen Pan[1], Jingxuan Ge[1], Yan Wang[4], Mingyang Wang[1], Zhenxing Wu[1], Xingcai Zhang[7], Jiameng Yu[5], Xujun Zhang[1], Tianyue Wang[1], Lirong Wen[6], Guangning Yan[8], Yafeng Deng[2], Hui Shi[2], Chang-Yu Hsieh[1,*], Zhihui Jiang[3,4,5,*], Tingjun Hou[1,*]

[1]*College of Pharmaceutical Sciences, Zhejiang University, Hangzhou 310058, Zhejiang, China*

[2]*CarbonSilicon AI Technology Co., Ltd, Hangzhou 310018, Zhejiang, China*

[3]*Department of Pharmacy, General Hospital of Southern Theatre Command, Guangzhou 510010, Guangdong, China*

[4]*School of Pharmaceutical Sciences, Southern Medical University, Guangzhou 510515, Guangdong, China*

[5]*Graduate School, Guangzhou University of Chinese Medicine, Guangzhou 510006, Guangdong, China*

[6]*School of Pharmaceutical Sciences, Dali University, Dali 671003, Yunan, China*

[7]*Department of Materials Science and Engineering, Stanford University, Stanford, CA 94305, USA*

[8]*Department of Pathology, General Hospital of Southern Theatre Command, Guangzhou 510010, China*

[#]*Equivalent authors*

## Corresponding authors

**Tingjun Hou**

E-mail: tingjunhou@zju.edu.cn

**Zhihui Jiang**

E-mail: jandsphy@163.com

**Chang-Yu Hsieh**

E-mail: kimhsieh@zju.edu.cn

## Abstract

Large language models (LLMs) have shown remarkable advancements in chemistry and biomedical research, acting as versatile foundation models for various tasks. We introduce AMP-Designer, an LLM-based approach for swiftly designing novel antimicrobial peptides (AMPs) with desired properties. Within 11 days, AMP-Designer achieved the de novo design of 18 AMPs with broad-spectrum activity against Gram-negative bacteria. In vitro validation revealed a 94.4% success rate, with two candidates demonstrating exceptional antibacterial efficacy, minimal hemotoxicity, stability in human plasma, and low potential to induce resistance, as evidenced by significant bacterial load reduction in murine lung infection experiments. The entire process, from design to validation, concluded in 48 days. AMP-Designer excels in creating AMPs targeting specific strains despite limited data availability, with a top candidate displaying a minimum inhibitory concentration of 2.0 μg/ml against *Propionibacterium acnes*. Integrating advanced machine learning techniques, AMP-Designer demonstrates remarkable efficiency, paving the way for innovative solutions to antibiotic resistance.

## Introduction

Bacterial antimicrobial resistance (AMR) poses a major global threat to human health. It has been estimated that almost 4.95 million deaths were associated with bacterial AMR globally in 2019, including 1.27 million deaths attributable to bacterial AMR(*1*), which had been predicted to rise to 10 million deaths per year by 2050(*2*). Gram-negative bacteria are particularly troubling as they have developed resistance to most commonly used antibiotics. This alarming situation is further exacerbated by the fact that no new class of antibiotics that are effective against Gram-negative bacteria has passed the clinical stage since the introduction of quinolones in 1968(*3*).

Typically ranging from 10 to 50 amino acids in length, antimicrobial peptides (AMPs) are naturally produced by various organisms to combat invading microorganisms(*4*). Due to their remarkable diversity in structures and functions, along with their good efficacy and reduced likelihood of resistance development, AMPs have been considered as potential alternatives to conventional small molecule antibiotics. However, compared to small-molecule-based antibiotics, many AMPs suffer from serious limitations such as relatively lower antibacterial activity, uncertain toxicity profiles, and susceptibility to inactivation during production and transportation(*5-8*), impeding their widespread application. To address these challenges, current research focuses on developing novel AMP candidates with enhanced activity, reduced toxicity, and increased resistance to hydrolysis. These improvements should make AMPs as contending antibiotics.

In recent years, there has been a growing interest in leveraging computer-aided methods to expedite the design of antimicrobial peptides(*9*). Conventional methods for designing AMPs typically involve optimizing existing AMPs or using AMP prediction models(*10-18*) to exhaustively screen a large-scale peptide space. However, the vast size of the peptide sequence space, estimated to cover approximately $4.5 \times 10^{41}$ peptides with the sequence length up to 32 residues(*19*), poses a formidable challenge of discovering new AMPs. Moreover, despite substantial efforts, only about 10 AMPs have been approved by regulatory agencies(*20*), indicating that the therapeutically relevant AMPs are sparsely distributed across this vast sequence space. Consequently, there is an urgent need to develop novel approaches for more effective AMP design.

With the advancement of deep generative modeling techniques, new AMP design methods have emerged(*21, 22*), including those based on recurrent neural networks (RNN)(*23-25*), generative adversarial networks (GAN)(*19, 26-29*), and variational autoencoders (VAE)(*30-33*). Designing AMPs with multiple desired traits, such as high activity and low hemolysis, is a non-trivial task. In order to generate AMPs with these desired traits, existing approaches rely on the paradigm of supervising learning with few labeled data. As models only learn from experimentally validated and labeled AMPs, this leads to low novelty for generated AMP sequences. One strategy to sample from a wider peptide space and to improve novelty is to train generative models on a broader data set containing all kinds of natural peptides. But this naive strategy cannot guarantee a high probability of generating AMPs with the desired traits. This dilemma (novelty versus validity) is usually formulated as a problem of multi-condition generations. Conditional GANs(*34*) and conditional VAEs(*35*) have emerged as popular solutions to this challenge(*19*), but their reliance on labeled data for training has been an issue. The process of labeling the dataset using AMP predictors or assuming unlabeled data as negative samples can introduce biases and result in substantial prediction errors during the training stage. Fine-tuning on small datasets of experimentally labeled AMPs is another feasible approach(*36, 37*). However, traditional methods of fine-tuning large-scale models suffer from high computational costs and the risk of overfitting to small datasets(*38-40*). Alternatively, a classifier trained on the latent space(*32, 41*) can identify regions where AMPs with desired traits reside and be combined with rejection sampling to generate the desired sequences with improved success rate.

The emergence of ChatGPT recently has captured the wildest imagination of the crowd. AMP generation is a specific task of protein language modeling, which has greatly benefited from the development of large language models (LLMs) for natural language processing (NLP). Drawing analogy from NLP and other protein design cases, we have reasons to believe that large-scale sequence generation models can also excel for AMP design. Adapting an LLM for downstream tasks, prompt tuning has gradually replaced traditional fine-tuning methods. Compared to fine-tuning, prompt tuning offers higher computational efficiency and greater flexibility without having to re-train the model. Moreover, it provides a means to mitigate overfitting to some extent and effectively bridges the gap between pre-training and downstream tasks(*39, 42, 43*). In the context of conditional generation tasks, achieving a balance between generation diversity and success rate is crucial, making prompt tuning an appropriate approach.

In this work, we proposed AMP-Designer, a comprehensive framework for AMP design that integrates GPT(*44*), prompt tuning(*45*), contrastive learning, knowledge distillation, and reinforcement learning (RL), followed by a series of wet-lab analyses for the validation of the designed novel AMPs (**Fig. 1**). Our approach involves training an AMP-centric language model as the foundation model, AMP-GPT, on a dataset of peptides extracted from UniProt(*46*). To facilitate transfer learning on a labeled dataset of AMPs and enable the generation of peptide sequences with desirable traits, we employed AMP-Prompt, *i.e.,* contrastive prompt learning while keeping the parameters of AMP-GPT frozen. To further reduce computational costs, we performed model distillation on AMP-Prompt, by compressing it into an RNN composed of three GRU layers, namely AMP-Distillation. Additionally, we constructed the minimum inhibitory concentration (MIC) prediction models, AMP-MIC, based on AMP-GPT for different bacterial species, providing feedback and facilitating subsequent screening with RL.

Utilizing AMP-Designer, we achieved rapid *de novo* design of highly diverse and potent AMPs against given bacteria species. Selection and synthesis of the top 20 AMP candidates (two peptides failed after three rounds of chemical syntheses) based on the average prediction scores provided by AMP-MIC enabled the discovery of two wet-lab validated exceptional AMPs, KW13 and AI18, within 48 days **(Fig. 1a)**. These two peptides exhibited strong *in vitro* antibacterial activity against a diverse set of Gram-negative bacteria and Gram-positive bacteria. Notably, both newly designed peptides demonstrated low off-target hemolysis toxicity, excellent plasm stability and strong antibacterial activity against a collection of clinically derived resistant Gram-negative bacteria. KW13 and AI18 at sub-MIC concentration did not induce drug resistance in *E. coli* after 30 generations of culture. Furthermore, they showed remarkable therapeutic efficacy in a bacterial pneumonia mouse model.

AMP-Designer is a plug-and-play framework. For new design tasks, it only takes around 3 days to complete the design based on the trained foundation model AMP-GPT. Notably, AMP-Designer demonstrates excellent efficacy in designing specific AMPs even in scenarios with severely restricted labeled data, which are frequently encountered in real-world situations. To illustrate this capability, we designed AMPs against *Propionibacterium acnes* (*P. acnes,* also known as *Cutibacterium acnes*), an anaerobic microorganism with extremely scarce labeled data points (less than 20). Given the impracticality of training an MIC predictor of *P. acnes* for guiding model optimization through RL in this context, we employed AMP-Designer to generate five AMP candidates against *P. acnes*, among which three displayed high potency during subsequent experimental evaluations. The successful few-shot AMP design highlights the commendable adeptness of the AMP foundation language model in conjunction with the antimicrobial peptide discovery workflow, in effectively navigating the antimicrobial peptide space.

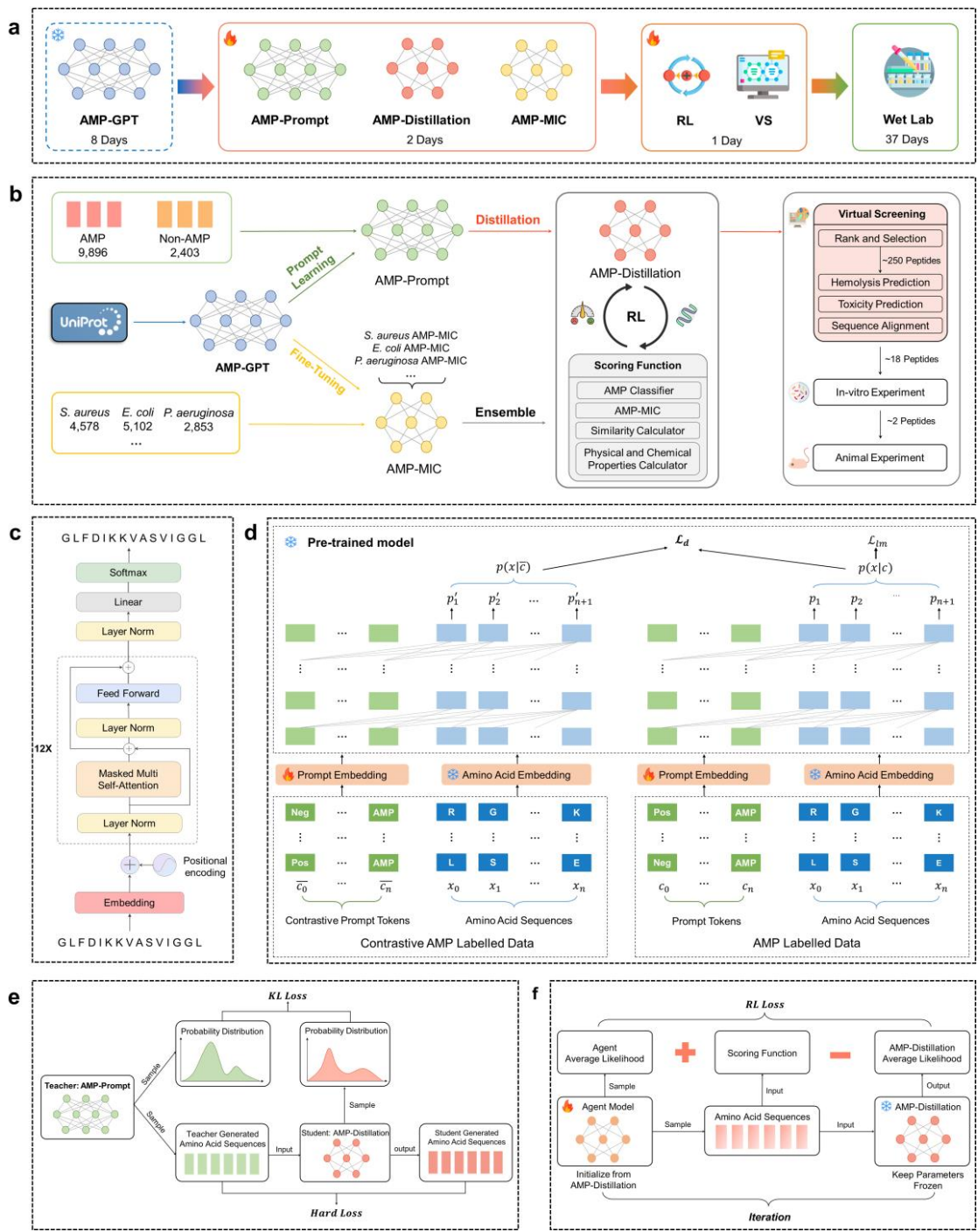

**Fig.1 | The overview of AMP-Designer. (a)**. The timeline for AMP screening using AMP-Designer, followed by the *in vitro* and *in vivo* validations; **(b).** the workflow of AMP-Designer; **(c).** Architecture of AMP-GPT; **(d).** Contrastive prompt fine-tuning process; **(e).** Model distillation process; **(f).** Reinforcement learning optimization process.

## Results

**AMP-Designer: the overview**

The AMP-Designer workflow is presented in **Fig. 1b**, depicting the model and training details. Initially, a collection of peptides with lengths less than or equal to 32 was gathered from UniProt(*46*), resulting in a dataset comprising 630,683 peptides. These peptides were used to conduct unsupervised training of AMP-GPT. Compared to training on small labeled datasets, leveraging a larger and more diverse dataset enables AMP-GPT to learn a more comprehensive representation of the data, a well-established phenomenon for large-scale language and biological models(*47-53*). Subsequently, the objective is to identify sequences with antimicrobial activity within the peptide space learnt by AMP-Designer. To achieve this, a curated set of 9,896 positive samples and 2,403 negative samples were retrieved from various antimicrobial peptide databases. Employing these labeled samples, contrastive prompt tuning was applied to fine-tune the model. Specifically, Initializing the prompt embedding layer based on the word embedding of AMP-GPT, and the labeled data was utilized to train this layer while holding the parameters of AMP-GPT fixed. The resulting model obtained through this process is referred to as AMP-Prompt.

While the integration of top-*k*(*54, 55*) sampling has shown remarkable success rates in generating AMPs (detailed in the section **Activity analysis**, where three AMP predictors were employed to assess the generated sequences), it is important to acknowledge that AMP-Prompt is prompt-tuned on a binary-class dataset of AMPs. Therefore, the model is limited to binary-class generation, lacking the ability to impose constraints on generating other desired properties. In a real-world scenario, a therapeutic AMP has to satisfy multiple requirements, such as manifesting higher activity or targeting specific bacteria. Therefore, further optimization is needed on top of AMP-Prompt to meet these additional criteria. ChatGPT has achieved tremendous success by leveraging the technique of reinforcement learning from human feedback (RLHF), where the core idea is to utilize expert knowledge to score the generated sequences and provide feedback for the RL agent. For AMP design, scoring the generated sequences based on human expertise is challenging (refer to the section **Methods** for more algorithm details).

Numerous properties influence the activity of AMPs. For instance, AMPs often exhibit a certain degree of hydrophobicity and hydrophilicity, contributing to their interactions with bacterial membranes. The balance between hydrophobic and hydrophilic characteristics can impact the targeting efficiency and solubilization capacity of AMPs. Additionally, the presence of a higher positive charge affects the charged nature of antimicrobial peptides, influencing their interactions with bacterial membranes. Cationic antimicrobial peptides, in particular, demonstrate an inclination to interact with anionic molecules within bacterial membranes, thereby eliciting perturbation of the cellular membrane. Furthermore, the prevalence of α-helical secondary structures is noteworthy. Many antimicrobial peptides adopt α-helical conformations during their interactions with membranes. This particular structural motif facilitates the insertion of antimicrobial peptides into the lipid bilayer of cellular membranes, thereby inducing membrane disruption. Therefore, building upon AMP-GPT, we trained the AMP-MIC predictors for different bacteria, considering properties such as electric charge, to provide feedback for RL (refer to the section **Design high-activity against Gram-negative bacteria peptides** for more details).

However, performing RL optimization directly on AMP-Prompt, which is based on GPT, requires substantial computational resources. To address this issue, we employed knowledge distillation, leading to the development of a more compact model termed AMP-Distillation that maintains a comparable generative capability to that of AMP-GPT. Subsequently, RL was applied to optimize AMP-Distillation, facilitating the identification and selection of the most optimal candidate AMPs suitable for the subsequent syntheses.

**Analysis of physical and chemical properties**

We randomly selected 2,000 peptide sequences from both the generative model and real AMPs for analysis. We employed an analysis tool from modlAMP(*56*) to visualize the distribution of amino acid frequencies, total charge, Eisenberg hydrophobicity, Eisenberg hydrophobic moment(*57*), and sequence length of these peptides. These fundamental attributes serve to facilitate our analysis of the model's learning dynamics, encompassing an assessment of whether the foundational model has effectively captured the distribution across the entirety of the small peptide space. Furthermore, this enables an evaluation of whether the subsequent fine-tuning of the model has effectively assimilated the distribution inherent to the real AMPs.

As shown in the **Fig. 2**, the distributions of the physicochemical properties of the peptides generated by AMP-GPT are similar to those of the UniProt training dataset. This observation demonstrates that AMP-GPT has effectively learned the physiochemical characteristics of peptides in the UniProt training dataset. After contrastive prompt learning, we observed a notable shift in the distribution of various physicochemical properties of the generated peptide sequences. The properties of these newly generated peptides were closer to those of the real AMPs, indicating the effectiveness of our prompt tuning. We further employed top-*k* sampling to generate peptides after prompt learning and found that, compared to traditional temperature-based sampling, top-*k* sampling resulted in a distribution of peptides with higher global charge and modified physicochemical properties, such as Eisenberg hydrophobicity and hydrophobic moment, that were more similar to those of the real AMPs. As most reported AMPs are cationic and amphiphilic in nature and possess properties that are thought to be crucial for insertion into and disruption of the bacterial outer-membrane which is the main cause of resistance and inactivity of many antibiotics[5]. Finally, the Distillation model produced peptide sequences with physicochemical property distributions that were highly similar to those generated by the teacher model, demonstrating that it successfully learned the probability distribution of the teacher model.

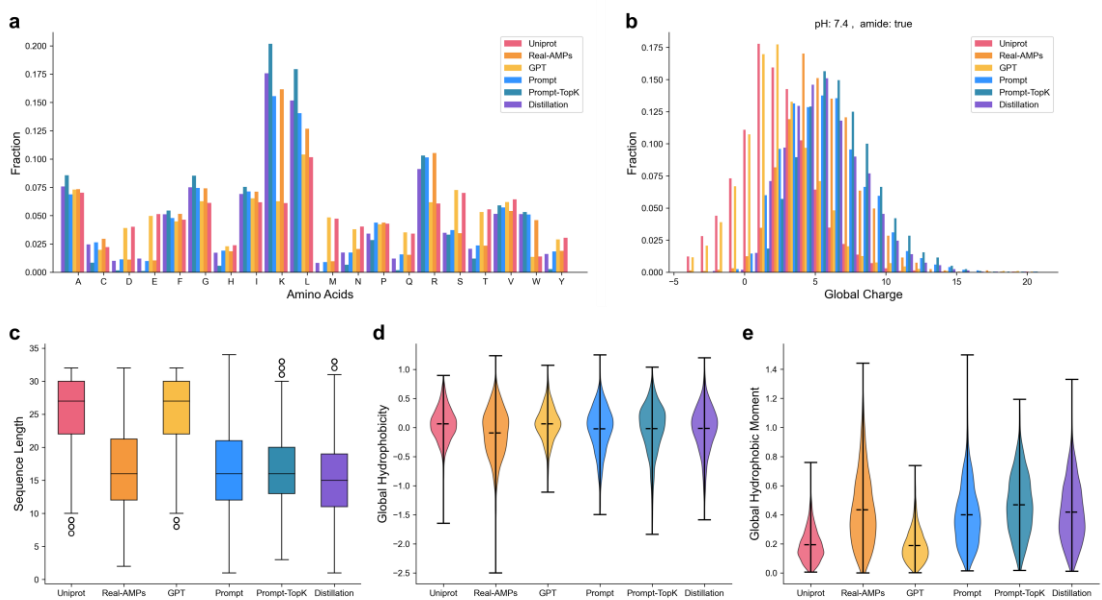

**Fig. 2 | Comparison of physicochemical properties.** UniProt (red), real AMPs (orange), AMP-GPT (yellow), AMP-Prompt (blue), AMP-Prompt-TopK (green) and AMP-Distillation (purple) distributions of the physical and chemical properties of peptides: **(a)** amino acid, **(b)** global charge, **(c)** sequence length, **(d)** Eisenberg hydrophobicity, and **(e)** Eisenberg hydrophobic moment.

**Activity analysis**

To evaluate the generated peptides, we utilized three different AMP classifiers: CAMP(*58*), AMP Scanner(*59*), and Macrel(*60*), to assess their potential for bioactivity. We introduced the state-of-the-art (SOTA) methods, CLaSS[32] and CFPS(*61*), as baselines for comparison. Following the recommendation by AMP Scanner that the predictions for sequences with less than 10 amino acids should be interpreted with caution, we selected multi-peptide sequences with a length of 10 or higher for our evaluation.

**Fig. 3a-c** illustrate a notable enhancement in the performance of the Prompt-based model across all three predictors. Particularly, the peptides generated by Prompt-TopK exhibit a distribution of bioactivity that closely mirrors that of real AMPs. Generally, a predictor assigns a sequence as an active AMP when the output is 0.5 or higher. It is evident that the performance of the candidate AMPs generated by various models aligns

consistently with the expectations across all three predictors. Notably, AMP-GPT demonstrates a strong ability to learn from the peptide data in UniProt, as the generated peptides exhibit an activity probability distribution to that of the peptide sequences from UniProt, consistently below 0.5 across all three predictors. However, after post-prompt tuning, the candidate AMPs generated by the model display activity probabilities comparable to those of real AMPs. Particularly, the performance of the Prompt-TopK variant even surpasses that of real AMPs, with probabilities centered around 1. Furthermore, our method consistently exhibits significantly higher probabilities of being predicted as active by all three predictors compared to CLaSS and CFPS.

Hence, we further calculated the percentage of the peptide sequences identified as AMPs by all three predictors simultaneously (with predicted values of 0.5 or higher). As illustrated in **Table 1**, the percentages are 50.7% for CLaSS and 57.2% for CFPS. In contrast, our approach achieves a substantially higher percentage of 83.4%, outperforming CFPS by 26.2%, marking a considerable advancement. Remarkably, our findings reveal that the predicted activity probability for peptides generated by Prompt-TopK exceeds that of real AMPs.

Furthermore, the Distillation model maintained nearly identical performance to the AMP-Prompt as shown in **Table S1**. Drawing upon the analysis of the physicochemical properties presented in the preceding section, it can be deduced that the student model, denoted as AMP-Distillation, has proficiently acquired the probability distribution inherent to the teacher model, AMP-Prompt.

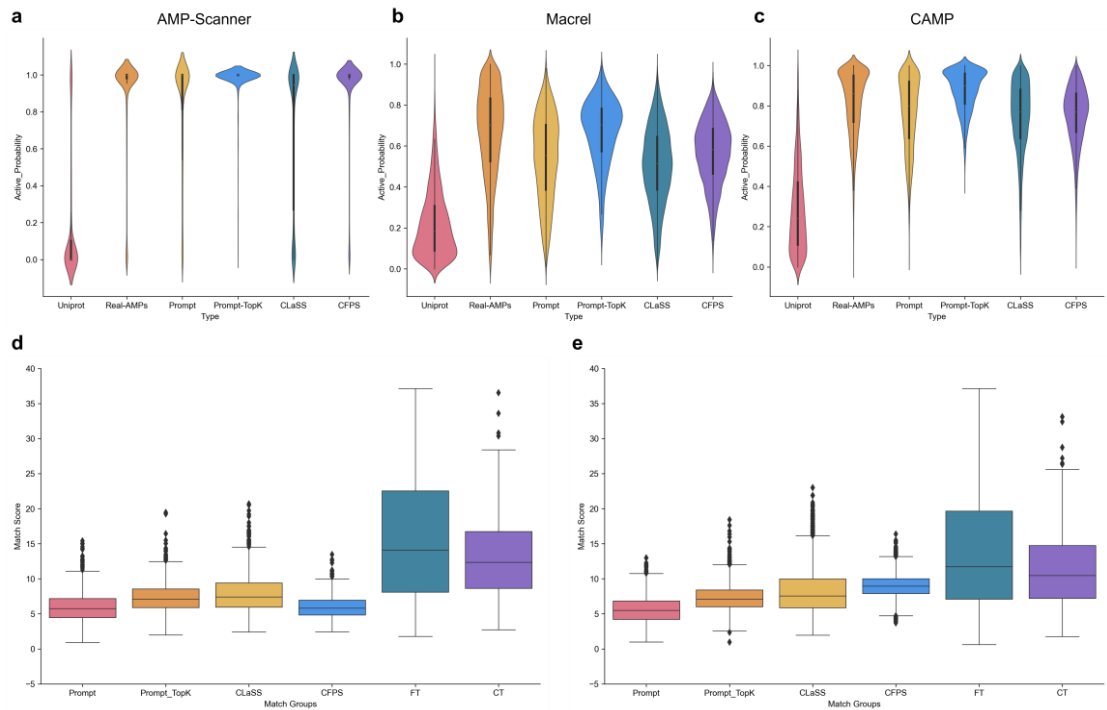

**Fig. 3 | Activity probability and match scores distribution. (a-c).** Activity probability distribution predicted by three AMP predictors. **(d-e).** Match scores distribution. Distributions of the match scores between **(d)** real AMPs and **(e)** generated internal sequences.

**Table 1 | Percentage of the peptides identified as AMPs by all three predictors simultaneously.**

| Type | Percentage |
| --- | --- |
| UniProt | 4.58% |
| Prompt | 61.8% |
| Prompt-TopK | **83.4%** |
| CLaSS | 50.7% |
| CFPS | 57.2% |
| Real-AMPs | 81.4% |

**Diversity and novelty analysis**

We further conducted an analysis on the Validity, Uniqueness, and Novelty of the peptides generated by our proposed method, as well as two baseline methods: Conditional Token (CT) and full model Fine-Tuning (FT). Specifically, the condition token method is retrained on labeled data and a property label is added to the model during training, which is used as the input during generation to autoregressively generate the entire peptide sequence. The traditional fine-tuning directly fine-tuned the AMP-GPT model using the real AMPs dataset. For detailed information, please refer to the **CT and FT models** section in the **Methods**. Validity indicates whether the generated sequence is valid, Uniqueness denotes the percentage of unique peptides generated, and Novelty represents the percentage of peptides that are different from the real AMPs.

In the analysis, it is evident that the CT produces invalid sequences, such as '[mask]', indicating a suboptimal learning of the distribution of amino acid sequences in AMPs. Furthermore, the comparison presented in **Table 2** reveals that the Uniqueness and Novelty metrics of the sequences generated by the FT are inferior to those generated by the prompt-based method. This observation suggests a higher degree of overfitting in the FT, as it generates sequences that exhibit repetition either within themselves or to the real AMPs.

This can be attributed to the fact that the CT and FT methods re-train the entire model on the real AMPs, which can easily cause overfitting. Conversely, our prompt-based approach freezes the pre-trained model parameters and only train the parameters of the embedding layer on the real AMPs, so they perform better than the CT and FT in terms of diversity and novelty, while maintaining a high activity likelihood for the generated peptides.

**Table 2 | Validity, Uniqueness and Novelty of five models.**

| Model | Validity | Uniqueness | Novelty |
| --- | --- | --- | --- |
| Condition Token (CT) | 97.9% | 91.8% | 84.4% |
| Fine-Tuning (FT) | 100% | 88.3% | 73.1% |
| AMP-Prompt | 100% | 99.9% | 99.6% |
| AMP-Prompt-TopK | 100% | 100% | 99.9% |
| AMP-Distillation | 100% | 99.8% | 99.7% |

To further explore the diversity of the generated peptides, we used the pairwise2.align.globalxx method from biopython(*62*) for sequence alignment and calculated the match scores of the generated sequences. The approach involves pairwise comparisons of the generated AMP sequences with the corresponding sequences (real AMPs or internal sequences) to determine the highest match score. **Fig. 3d** illustrates the match scores between the generated sequences and the real AMPs. Notably, our approach illustrates a lower resemblance to the real AMPs compared to the baselines. Additionally, we compared the internal similarity among the generated AMP sequences across various models (**Fig. 3e**), revealing that our method demonstrates overall lower internal similarity relative to the other models. The findings suggest that our approach not only enhances the probability of activity but also maintains a higher degree of diversity, leading to the generation of more novel peptide sequences.

**Design high-activity against Gram-negative bacteria peptides**

While the outlined approaches yield bioactive candidate AMPs, quantifiable control over the inherent properties of the generated candidates, such as their degree of activity and net charge, remains elusive. Therefore, in order to obtain AMP with higher activity and other desired properties, RL was employed for further optimization.

In the course of RL optimization, the reward function incorporates five distinct components, namely Classification Probability (*CP*), Minimum Inhibitory Concentration (MIC) against three bacterial strains, Sequence Length (*SL*), Memory Similarity (*MS*), and Charge. Detailed information regarding these components can be found in **Table S4**. The classification probability is the prediction result calculated by Macrel, which is a state-of-the-art (SOTA) AMP predictor(*63*). Meanwhile, the MIC was predicted by three AMP-MIC models which were trained using the MIC data of three different bacteria (*S. aureus*, *P. aeruginosa* and *E. coli*). The accuracy of the AMP-

MIC predictor is compared to that of the RNN model proposed by Huang et al.(*64*) and the GPT-based model trained from scratch, as shown in **Table S5**. Our experimental results demonstrate that the AMP-MIC achieves a higher accuracy than the other two models. To increase the diversity of the generated sequences, we did not directly use the continuous numerical values of the classification probably and MIC as the rewards in RL.

To clarify the contributions of RL, we conducted a comprehensive analysis by sampling molecules at each step during the RL tuning process. This allowed us to explore the relationship between the number of RL steps and the average scores achieved for generated AMPs. By examining the variations in average scores at different RL steps, we gained insights into the impact of RL on the generation of AMPs and the corresponding improvements in their quality.

As shown in **Fig. S3a-c**, the x-axis represents the RL steps, while the y-axis represents the respective average scores (AMP predicted score, *E. coli* predicted MIC and *P. aeruginosa* predicted MIC). It is evident that the average scores (activity probabilities) of the AMP predictor increase from 0.5 to over 0.7 as the number of steps in RL increases. Furthermore, the predicted MIC against *E. coli* and *P. aeruginosa* decreases from over 500 to below 5.

Subsequently, we visualized the distributions of three properties for the generated AMPs at 10, 30, and 50 steps. As depicted in **Fig. S3d**, there is a clear improvement in the average scores of the AMP predictor from the 10th step to the 50th step. To better visualize the distributions of predicted activity against *E. coli* and *P. aeruginosa*, we capped values above 20 at 20. As shown in **Fig. S3e-f**, the distribution of activity becomes increasingly concentrated from the 10th step to the 50th step, with most values falling below 5.

Moreover, we employed a memory queue to store the top 256 AMPs with the highest scores in order to increase the diversity of the generated sequences. During the RL process, 128 amino acid sequences were sampled to evaluate the current state, and those with higher scores were added to the memory queue, replacing those with lower scores. Memory similarity was calculated by comparing the sampled peptides with the peptides in the memory queue using the biopython's pairwise2.align.globalxx method(*62*), and then divided by the length of the sequence. The final reward, which incoporates all the aforementioned factors, is given in Equation 18. We sorted the scores of the generated peptide sequences and retained the top 250 peptides produced by the RL agent for further screening. Next, we calculated the AMP classification probabilities for CAMP, AMP Scanner and Macrel, the α-helical score calculated by biopython, the hemolysis score predicted by HemoPI(*65*), and the toxicity score predicted by ToxinPred(*66, 67*). Based on the aforementioned screening procedure, we retained the predicted AMP sequences that exhibited an α-helical structure with a hemolytic probability below 0.6 and were predicted to be non-toxic. Subsequently, we sorted these sequences based on their predicted scores for *P. aeruginosa* and *E. coli*. After excluding all non-compliant candidates, we identified the top 20 candidates based on the average prediction scores given by AMP-MIC testing against two Gram-negative bacteria: *P. aeruginosa* and *E. coli*.

**In vitro activities of the leading AMPs against members of ESKAPE**

The top 20 predicted AMPs were chosen for chemical synthesis and verification. Two peptides failed to synthesize even after three rounds of attempts. We then tested *in vitro* antibacterial activities of the 18 candidates. Firstly, we determined the MIC values of those AMPs against six standard strains of all members of ESKAPE (*E. faecalis* ATCC 29212, *S. aureus* ATCC 29213, *K. pneumoniae* ATCC 700603, *A. baumannii* ATCC 19606, *P. aeruginosa* ATCC 27853 and *E. coli* ATCC 25922). As shown in **Table 3**, 17 out of 18 candidates exert notable antibacterial activity against at least one strain. In general, the majority of AMPs are more effective against Gram-negative bacteria than Gram-positive bacteria, which is to some extent aligned with our specific design of AMPs targeting Gram-negative bacteria. Moreover, the remarkable activity of these AMPs, initially designed for *P. aeruginosa* and *E. coli*, extends to *K. pneumoniae* and *A. baumannii* as well, suggesting that these AMPs may possess a general bactericidal mechanism against Gram-negative bacteria. AI18, KW13, KW20, RV15 and GI16 exhibited the highest antibacterial activity against both Gram-negative bacteria (MICs range from 4 to 16 μg/ml) and Gram-positive bacteria (MICs range from 8 to 32 μg/ml), suggesting more excellent broad-spectrum activities across all tested ESKAPE strains. Consequently, we further evaluated the activities of AI18, KW13, KW20, RV15 and GI16 against six strains of clinically resistant Gram-negative bacteria (PDR *K. pneumoniae* 418015, XDR *K. pneumoniae* 325016, MDR *K. pneumoniae* 327004, MDR *E. coli* 103231, MDR *P. aeruginosa* 304238 and MDR *A. baumannii* 316039), and the results are shown in **Table 4**. All five AMPs showed notable activities against five strains of superbugs, with MICs ranging from 4 to 16 μg/ml; however, it is worth noting that KW13 and AI18 demonstrated only moderate activities against PDR *K. pneumoniae* 418015, a colistin-resistant strain (**Table S6**).

A thorough analysis was performed to compare the MIC values and sequence diversity of the AMPs designed by AI in recent studies(*32, 64*). The sequence diversity was assessed between the generated AMPs and biological patent sequences using the patseqfinder tool available on LENS.ORG(*68*). The outstanding performance of our designed peptides across various aspects is evident from the results presented in **Table S7**.

**Table 3 |** *In vitro* **antibacterial activities of top 18 AMPs against six standard ESKAPE strains.**

| AMPs | MIC[a] (μg/ml) | | | | | |
|---|---|---|---|---|---|---|
|  | ATCC 29212[b] | ATCC 29213[c] | ATCC 700603[d] | ATCC 19606[e] | ATCC 27853[f] | ATCC 25922[g] |
| RR13 | 128 | 64 | 64 | 32 | 32 | 16 |
| AI18 | 32 | 32 | 16 | 8 | 8 | 8 |
| KW20 | 16 | 16 | 16 | 16 | 8 | 4 |
| KW13 | 16 | 8 | 8 | 8 | 4 | 4 |
| VK18 | 64 | 16 | 16 | 8 | 16 | 16 |
| SK18 | 16 | 8 | 32 | 8 | 16 | 8 |
| RV15 | 8 | 8 | 16 | 8 | 16 | 8 |

| ID | | | | | | |
|---|---|---|---|---|---|---|
| IR20 | 32 | 32 | 32 | 16 | 64 | 32 |
| RK17 | 64 | 32 | 32 | 8 | 16 | 32 |
| LW14 | 16 | 16 | 32 | 16 | 64 | 16 |
| GI16 | 32 | 8 | 16 | 8 | 16 | 16 |
| RR16 | 256 | 64 | 16 | 16 | 4 | 8 |
| YK19 | 16 | 16 | 64 | 32 | 32 | 32 |
| AK15 | 64 | 8 | 16 | 16 | 8 | 8 |
| KK12 | >256 | >256 | >256 | >256 | >256 | >256 |
| KWI18 | 64 | 32 | 64 | 16 | 64 | 32 |
| LF19 | 16 | 8 | 32 | 8 | 16 | 16 |
| AV20 | 16 | 8 | 16 | 8 | 32 | 8 |

[a] Minimal inhibitory concentration; [b] *E. faecalis*; [c] *S. aureus*; [d] *K. pneumoniae*; [e] *A. baumannii*; [f] *P. aeruginosa*; [g] *E. coli*.

**Table 4 | *In vitro* antibacterial activities of four selected AMPs against six resistant clinically Gram-negative strains.**

| ID | MIC (μg/ml) | | | | | |
|---|---|---|---|---|---|---|
| | PDR *K. pneumoniae* 418015 | XDR *K. pneumoniae* 325016 | MDR *K. pneumoniae* 327004 | MDR *E. coli* 103231 | MDR *P. aeruginosa* 304238 | MDR *A. baumannii* 316039 |
| AI18 | 32 | 8 | 16 | 4 | 16 | 8 |
| RV15 | 16 | 16 | 16 | 8 | 16 | 8 |
| KW13 | 32 | 8 | 16 | 8 | 8 | 4 |
| KW20 | 16 | 8 | 16 | 4 | 8 | 8 |
| GI16 | 16 | 16 | 16 | 4 | 32 | 4 |

Off-target toxicity, especially hemolysis, is a major concern for cationic AMPs. We evaluated hemolysis of human erythrocytes in the presence of AI18, KW13, KW20, RV15 or GI16. As displayed in **Fig. 4a**, KW13 and AI18 demonstrated negligible hemolysis effects. We further determined the 50% hemolytic concentration (HC50, 50% hemolysis) of these two AMPs. Their HC50 values were both higher than 1000 μg/ml, which quantitatively demonstrated their excellent hemocompatibilities (**Fig. S4**).

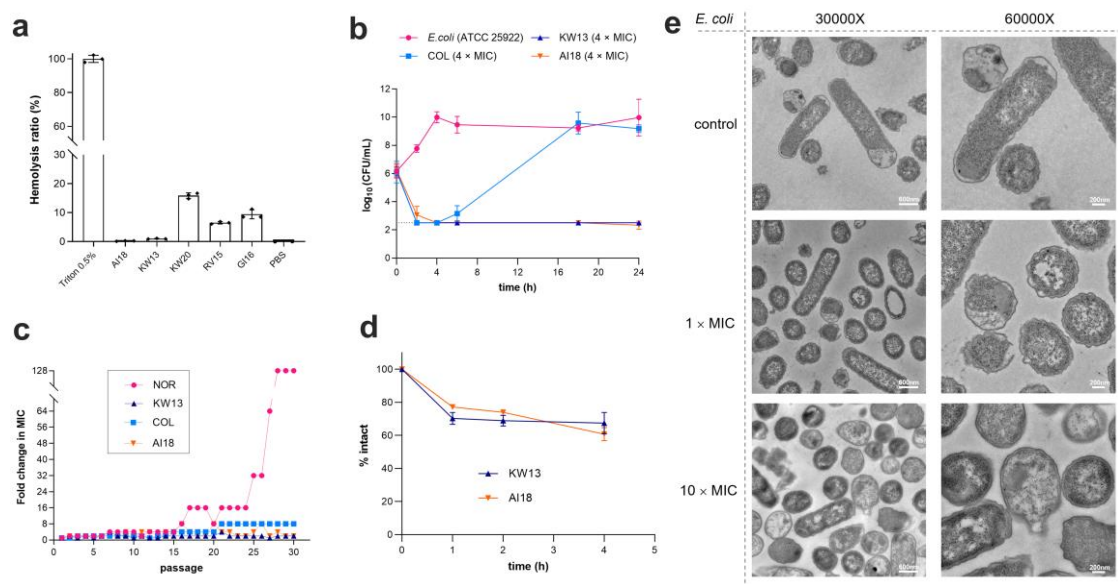

**Fig. 4 | Biological properties of the selected AMPs *in vitro*.** (**a**). Hemolytic activity of KW13, AI18, KW20, RV15 and GI16 at a concentration of 100 μg/ml (n = 3 biologically independent replicates, mean ± s.d.). (**b**) Time–kill curve of KW13 and AI18, with colistin (COL) as positive control. (n = 3 biologically independent replicates, mean ± s.d.). (**c**). Resistance induction studies of *E. coli* ATCC 25922 when cultured in the presence of sub-MIC (1/2×) levels of KW13 and AI18, with norfloxacin (NOR) and COL as positive control. (**d**). Human plasma stability testing of KW13 and AI18 by high performance liquid chromatography (HPLC). Data are shown as mean ± s.d. (n=3 biologically independent). (**e**). TEM images of *E. coli* cells treated with KW13 and PBS (control) at 30000× magnification (scale bar, 600 nm) and 60000× magnification (scale bar, 200 nm).

To more comprehensively assess the bactericidal effects of KW13 and AI18, we conducted time-kill studies against *E. coli* ATCC 25922. As shown in **Fig. 4b**, KW13 and AI18 can significantly reduce viable counts after 2 h incubation, and the bactericidal effect lasted at least for 24 h, while the last resort antibiotic colistin only exerts bactericidal effect for about 8 h.

The propensity to induce bacterial resistance is recognized as a challenging feat for AMPs, given their mode of action which primarily targets bacterial membranes rather than engaging a specific receptor(*69*). We assessed possible antibacterial resistance by culturing *E. coli* ATCC 25922 with sub-MIC level of KW13 and AI18, using the classic small-molecule antibiotic norfloxacin for comparison. As shown in **Fig. 4c**, *E. coli* could not develop obvious resistance against KW13 and AI18 after 30 generations, as is the case against colistin. In contrast, the bacteria started to develop resistance to norfloxacin after 17 generations, with a 128-fold reduced susceptibility after 30 generations.

As the plasma stability of an AMP is an important factor in its ability to exert *in vivo* efficacy, we first evaluated the proteolytic stability of KW13 and AI18 before evaluating their *in vivo* antibacterial efficacy. As shown in **Fig. 4d**, KW13 and AI18 degraded slowly in human plasma in 4 h, indicating that KW13 and AI18 possess greate plasma stability.

We then focused on elucidating the membrane disruption effect of KW13, which showed the highest activity *in vitro* against *E. coli*. The common mechanism of action of AMP is bacterial lysis by forming pores on bacteria membranes(*70*). We used transmission electron microscopy (TEM) to observe the potential morphological changes of membrane in *E. coli* ATCC 25922 cells upon KW13 treatment (MIC = 4 μg/ml, treated with 1× and 10× MIC for 5 h). **Fig. 4e** demonstrated that the cell content leaked out and cell lysed at 1× and 10× MIC in a dose-dependent manner, suggesting that the integrity of the bacteria cell membrane decreased.

Taken together, the *in vitro* results indicate that KW13 and AI18 passed a preliminary safety profiling and hold therapeutic potential for treating Gram-negative bacteria. Therefore, we conducted further *in vivo* evaluation.

***In vivo* activities of KW13 and AI18 against bacterial pneumonia in a mouse model**
We then tested the therapeutic efficacy of KW13 and AI18 by constructing an XDR *K.*

*pneumoniae* 325016 infected pneumonia mouse model. Hospital-acquired pneumonia infected by carbapenem-resistant *Klebsiella pneumoniae* (CRKP) is a major clinical concern, which is difficult to be treated and therefore is associated with high mortality(*71*). AMP has been recognized as a promising antibiotic alternative to treat lung infections. The experimental setup is shown in **Fig. 5a.** The bacterial pneumonia model was constructed by suppressing mice immunity with cyclophosphamide, followed by intratracheal administration of *K. pneumoniae* 325016. Two doses of AMPs were given and the lungs were obtained after treatment to count the bacteria load. As shown in **Fig. 5b,** two AMP-treated groups presented reductions of ~99% in bacteria load, which was significantly lower than that in the blank control PBS group, and their therapeutic effects were comparable to that of the positive control AMP indolicidin(*72, 73*). Tissue histological analysis showed that the lungs treated with KW13, AI18 and indolicidin remained healthy and had clear alveoli morphology, while the PBS group showed severe tissue damage due to the bacterial infection (**Fig. 5c**).

The above results demonstrate that KW13 and AI18 exert promising *in vivo* antibacterial activities against resistant *K. pneumoniae* infected pneumonia without any obvious adverse effects on the host and deserve further investigations.

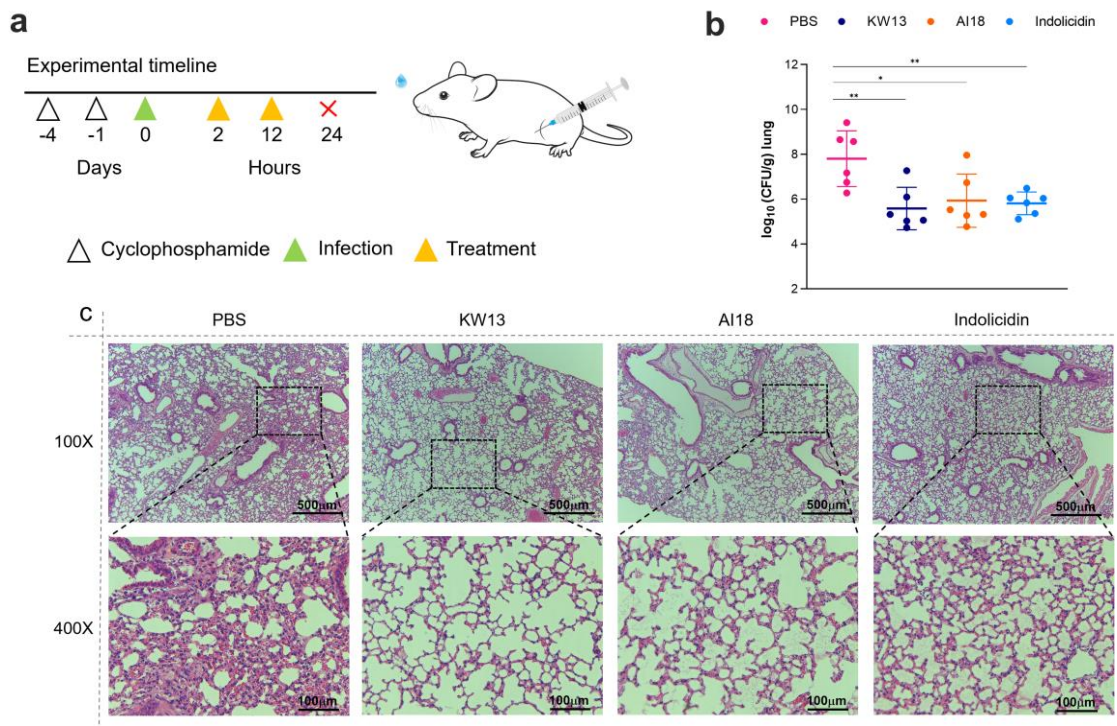

**Fig. 5 | *In vivo* efficacy of KW13 and AI18 in treating mouse pneumonia. (a)**. Mouse infection model construction procedures. Mice were injected with cyclophosphamide intraperitoneally twice, followed by the tracheal instillation of XDR *K. pneumoniae* 325016 to establish the pneumonia model. The AMPs (10 mg/kg) were injected intraperitoneally twice, and the mice were killed 24 h after infection. **(b)**. Bacteria load in the lungs of infected mice after treatment with PBS, KW13, AI18, and indolicidin (n = 6 biologically independent replicates, mean ± s.d.). Statistical analysis was performed using one-way ANOVA. **(c)**. Histological images of the lung tissue sections. First row: an overview of the lung tissues after H&E staining (scale bar, 500 μm). Second row: detailed histological images in the regions of black boxes (scale bar, 100 μm). Experiments were performed in duplicates with similar results and one representative figure is shown.

**Few-shot AMP design against *P. acnes***

Acne vulgaris is a chronic skin disorder, which affected more than 630 million people globally, thus being ranked as the eighth most common disease in the world(*74*). Colonization of anaerobic Gram-positive bacterium *P. acnes* in the pilosebaceous follicles of human skin is closely related to the severity of acne vulgaris(*75*). Conventional small molecule antibiotics were commonly used topically or systemically based on the severity of the acne lesion. However, resistance to these antibiotics have been increasing reported(*76, 77*).

The challenge in this task stems from the scarcity of labeled data pertaining to *P. acnes*. Upon analysis of the entire fine-tuning dataset, a total of 47 AMPs is identified with labels indicating explicit activity against *P. acnes*. Among these peptides, 17 possess C-terminal amidation and exhibit a length equal to or less than 32 amino acids. Our analysis convincingly revealed a positive correlation between the MIC values against *P. acnes* and the antimicrobial activity against *S. aureus*, *Staphylococcus epidermidis* (*S. epidermidis*) and *P. aeruginosa*. In light of this correlation, we employed three predictors of antimicrobial activity against *S. aureus, S. epidermidis* and *P. aeruginosa* as the proxy for gauging the potency against *P. acne*. The overall process closely resembled the one employed in the "**Design high-activity against Gram-negative bacteria peptides**" section. Ultimately, we selected 5 candidate AMPs for subsequent *in vitro* experimental validation.

**Table 5|** ***In vitro*** **antibacterial activities of five AMPs against *P. acnes*, *S. aureus* and *E. coli***

| AMPs | MIC (µg/ml) | | | |
| --- | --- | --- | --- | --- |
| | *P. acnes* ATCC 6919 | *P. acnes* ATCC 11827 | *S. aureus* ATCC 29213 | *E. coli* ATCC 25922 |
| **FI19** | 2 | 4 | 128 | 64 |
| **IF20** | 4 | 8 | 64 | 64 |
| **RW20** | 64 | 8 | 128 | 64 |
| **RW12** | 32 | 64 | 64 | 64 |
| **KW18** | 64 | 16 | >256 | >256 |

As shown in **Table 5**, three out of the five designed AMPs showed potent activities against *P. acnes*, with MICs ranging from 2 to 8 µg/mL, which are the lowest compared to those against gram-positive bacteria *S. aureus* and gram-negative bacteria *E. coli*. The results indicate the specificity of these three AMPs against *P. acnes*, aligning perfectly with our intended design goal for a selectively targeted AMP. Among the three candidates, FI19 displayed the highest efficacy against *P. acnes* with an impressive MIC value of 2.0 µg/ml. Property-activity correlation analysis demonstrated that higher hydrophobicity combined with lower net charge likely contributes to enhanced antimicrobial activity. As potent, narrow-spectrum AMPs specifically targeting *P. acnes*, further evaluation and modification of FI19, IF20 and RW20 have great potential in transforming them into highly promising anti-acne antibiotics for systemic use.

## Discussion

Peptide design poses a highly challenging task, and unlike small molecules, peptides are composed of amino acid sequences and their structures exhibit high flexibility, making expert-based design a daunting process. In recent years, with the advancement of artificial intelligence (AI), an increasing number of AI-based approaches have been utilized for drug or material design. Given the similarities between the characteristics of peptides and natural language, leveraging the powerful capabilities demonstrated by recent advancements in GPT, it is believed that employing similar language models can expedite the discovery of peptides.

In this study, we present an AMP-centric foundation language model and innovative framework for AMP design, denoted as AMP-Designer. This framework includes the development of the first large-scale pre-trained generative model for AMPs, known as AMP-GPT. By incorporating prompt tuning, knowledge distillation, and reinforcement learning, we notably enhanced the efficiency of AMP generation. Particularly, the integration of prompt tuning resulted in an impressive activity prediction rate of 83.4% for AMPs using three SOTA AMP classification predictors, surpassing the proportion observed for real AMPs (81.4%), while maintaining higher sequence diversity. To further optimize the model for improved AMP activity, we employed RL based on multiple properties. Additionally, knowledge distillation was applied to reduce computational costs. To validate the performance of AMP-Designer, we specifically focused on designing antimicrobial peptides against Gram-negative bacteria. Subsequently, 18 peptides were selected for experimental validation.

The *in vitro* experiments confirmed the broad-spectrum effectiveness of the chosen AMPs, of which the top five most active showed potent antibacterial activities against a wide range of clinically relevant Gram-negative pathogens. KW13 and AI18, as two leading AMPs, exhibited therapeutic outcomes comparable to indolicidin in bacterial pneumonia mouse models, with little hemolysis toxicity and a low tendency to induce drug resistance. These results highlight the potential of our AMP-designer to tackle the pressing issue of the escalating AMR on a global scale and the dwindling antibiotic discovery pipeline. KW13 and AI18 can serve as starting points for structure optimization via peptide cyclization, stapling or fatty acid modification to support further potency increases and plasma stability enhancement. Additionally, to broaden the application of the two AMPs, research on their combination with antibiotics is also necessary.

Furthermore, we devised a few-shot design task targeting *P. acnes*. Leveraging the observed correlation between the AMPs against *P. acnes* and *S. aureus*, we effectively engineered AMPs with exceptional in vitro activity against *P. acnes*. While the fine-tuning dataset encompasses approximately 20 instances bearing experimental labels specific to *P. acnes*, the overall corpus of authentic real AMPs dataset comprises nearly 10,000 distinct AMPs. It is noteworthy that AMP-Designer does not explicitly focus on assimilating the information from these *P. acnes* targeting AMPs. Moreover, this limited subset is inadequate to adequately train an MIC predictor specialized for *P. acnes*, thereby impeding the capacity to optimize for the antimicrobial activity against *P. acnes* in the reinforcement learning process. Nonetheless, the framework manages to generate remarkably potent AMPs effective against *P. acnes*. This attests to the efficacy of the proposed AMP design workflow based on the peptide foundation model, which has adeptly captured the spatial distribution of diverse antimicrobial peptides. This capability remains salient even in scenarios where pertinent labeled data is scarce, exemplifying its ability to probabilistically generate highly active AMPs.

Looking ahead, we see an increasing role for cross-modalities of research comprising closely connected, mutually reinforcing experimentation and deep learning. Human-directed experimental trial-error-optimize search strategies of the huge peptide sequence space can frequently lead to highly inefficient resources deployment. Integrating deep learning with targeted wet experiment to guide efficient experimentation and provide new data to improve algorithm performance creates a mutually beneficial cycle that saves labor, money, and time to greatly accelerate AMP discovery and design.

Furthermore, we observed notable potential in generating multiplexed peptide activity through the integration of LLM with the optimization paradigm. The workflow described in this work are generic and extensible, based on the AMP pre-training model AMP-GPT, thereby enabling its seamless integration into various downstream peptide design tasks. The project, encompassing *in silico*, *in vitro* and *in vivo* aspects, spanned a total of 48 days, in which the computational phase took 11 days. Notably, leveraging the existing AMP-GPT foundation model is expected to further shorten this duration, enabling researchers to accomplish specific peptide generation tasks within a timeframe of 3 days. This permitting them to be straightforwardly translated to the generation of peptides with diverse functions, including immunomodulation, cellular transportation, signaling, metabolism process.

## Methods

### AMP-GPT

AMP-GPT is built based on GPT2 small, consisting of 12 layers of Transformer(*78*) decoder and 8 attention heads. AMP-GPT outputs the token of each amino acid in the peptide in an autoregressive manner. The next generated token is determined by the previously generated character sequence, and masking is used to prevent information leakage from the unencoded characters in the sequence. The multi-head self-attention layer is the core of the model, composed of several scaled dot-product attention functions, which enables the model to sequentially capture key information. The attention mechanism is shown in Equation 1:

$$Attention(Q, K, V) = \mathrm{softmax}\left(\frac{QK^T}{\sqrt{d_k}}\right) V. \tag{1}$$

where $Q$, $K$, and $V$ represent the query, key, and value matrices, respectively, and $d_k$ is the dimension of $K$. Since the attention layer can process the entire sequence at the same time, additional positional encoding should be added to each token to ensure the model is aware of the linear sequence order.

To indicate the beginning or end of the sequence during sampling, it is necessary to define a start token and an end token, denoted as "CLS" and "SEP", respectively. During the training, the "CLS" token is concatenated to the amino acid sequence as the input, and the amino acid sequence concatenated with the "SEP" token is used as the label for model training.

The objective during the training phase is to minimize the negative log-likelihood, as shown in Equation 2:

$$\mathcal{L} = -\sum_{i=1}^{n} \log p\left(x_i | x_{<i}\right). \tag{2}$$

During the generation phase, peptides strings are generated using an autoregressive approach based on peptide fragments, which are then concatenated together as shown in Equation 3:

$$p(x) = \prod_{i=1}^{n} p(x_i | x_{<i}). \tag{3}$$

The generation process initiates with a start token, after which the model autoregressively generates amino acid tokens until reaching an end token.

**AMP-Prompt**

To enable the model to generate peptide sequences with desired properties, we need to perform transfer learning on the pre-trained peptide generation model AMP-GPT on a specified small dataset. Traditional language models tend to use a two-stage training approach, where they first pre-train on a large-scale corpus and then fine tune on the downstream task, which can be affected by various factors such as data annotation quality and overfitting. In addition, for different downstream tasks, the entire model needs to be fine-tuned, which can consume a lot of computing resources.

In this task, we need to ensure that the generated peptides have multiple properties and high diversity as well. To address this issue, inspired by the work of Jing et al., we use contrastive prompt tuning to perform transfer learning for the downstream task. As shown in **Fig. 1b**, this prompt method belongs to the technique of soft prompt(*38*), where during training, we keep the AMP-GPT parameters frozen, initialize a prefix embedding layer for positive word labels, and initialize another prefix embedding layer for negative word labels. This design enables the model to acquire a better sense of distinction between positive and negative samples using prompts.

The design of the loss function is crucial in contrastive learning. The loss function for the contrastive prompt training consists of two parts: the language modeling training loss and the contrastive discriminative loss. The language modeling loss is modified from Equation 2 by adding the prefix embedding as a condition. For each labeled sample $(x, c)$ where $x$ is the input sequence and $c$ is the label, the language modeling loss can be defined as:

$$\mathcal{L}_{lm} = -\sum_{i=1}^{n} \log p(x_i|x_{<i}, c). \tag{4}$$

The equation for the discriminative loss is as follows:

$$\mathcal{L}_d = -\log \frac{p(x|c)}{p(x|c) + p(x|\bar{c})}. \tag{5}$$

During training, the optimizer optimizes $\mathcal{L}_d$ to improve the attribution $p(c|y)$ by increasing $p(x|c)$ and lowering $p(x|\bar{c})$.

The final loss function is shown as Equation 6, and when $\omega_1 = 1$, there is no use of contrastive loss. At the same time, we used different sampling methods during generation: top-$k$ sampling and regular temperature sampling(*79*).

$$\mathcal{L}_{contrast} = \omega_1 \mathcal{L}_{lm} + \omega_2 \mathcal{L}_d. \tag{6}$$

**AMP- Distillation**

Using AMP-Prompt can improve the proportion of AMPs in the generated sequences, but our goal is to generate AMPs with increased activity as well as other desired properties. Therefore, our strategy involves optimizing the existing model through RL. However, directly fine-tuning RL on large models requires a lot of computation. To address this issue, we use knowledge distillation to distill the contrastive prompt-based AMP-Prompt with a smaller model.

Knowledge Distillation is a model compression technique that aims at training a smaller model to mimic a larger pre-trained model (or ensemble of models). This training setup is sometimes referred to as "teacher-student" with the large model as the teacher model and the small model as the student model. The method was first introduced by Bucila et al. in 2006(*80*).

The main idea is that certain knowledge regarding the dataset is transferred from a teacher model to a student model by minimizing the loss function. The goal of the student model is to learn the probability distribution predicted by the teacher model, which is the softmax layer's output of the teacher model.

In this work, we used a RNN as the student model to learn AMP-Prompt, which we referred to as AMP-Distillation. To enable AMP-Distillation to learn more comprehensive information, different from the method proposed by Wang et al.(*40*), we trained it using soft-target distillation. Specifically, we used the probabilities output by the softmax layer of AMP-Prompt as the labels for training. The target function $p_i$ can be defined as:

$$p_i = \frac{exp(z_i)}{\sum_j exp(z_j)}. \quad (7)$$

Through this approach, the AMP-Prompt model is transformed into an AMP-Distillation model that requires lower computational resources.

**Reinforcement learning**

RL is widely used for fine-tuning pre-trained neural networks. For example, RL is used to fine-tune RNNs for generating music with specific content(*81*). The success of this case demonstrates the feasibility of using RL for fine-tuning strategies in sequence models.

The goal of RL is to find an appropriate policy that maximizes the reward. In this work, we used the REINFORCE(*82*) algorithm to fine-tune the AMP-Distillation. REINFORCE is a policy gradient-based RL algorithm that uses Monte Carlo to determine the optimal policy. As shown in Equation 8, we use REINFORCE to optimize the pre-trained model parameter $\theta$ for the task of generating peptide sequences, so that the optimized model can generate peptides with desired properties and maximize the expected reward of each peptides sampling.

$$\theta' = argmax_\theta \left( E_{\pi_\theta}(G(\tau)) \right). \quad (8)$$

where $\pi_\theta$ represents the policy with the model parameter θ, and $\tau$ represents the set of state-action pairs $(s_t, a_t)$ in a trajectory from the initial state to the terminal state within $T - 1$ steps, as shown in Equation 9:

$$\tau = (s_0, a_0, \dots, s_{T-1}, a_{T-1}). \quad (9)$$

The reward for each trajectory to the $t$ step is $r(s_t, a_t)$. The sum of rewards from

the $t$ step to the final state is represented by Equation 10:

$$G(s_t, a_t) = \sum_{t=0}^{T-1} r(s_t, a_t). \tag{10}$$

According to the REINFORCE algorithm, the objective function can be derived as follows:

$$J(\theta) = E_{\pi_\theta}(G(s_t, a_t)) = \left(\sum_{t=0}^{T-1} \log \pi_\theta(s_t|a_t) G(s_t, a_t)\right). \tag{11}$$

In the task of generating peptide sequences, $G(s_t, a_t)$ at each step in each trajectory cannot be calculated (the score of the complete peptide cannot be estimated based on the fragment of the peptide). Therefore, this task is a sparse reward problem in RL. To apply the REINFORCE algorithm to this task, the goal of the complete peptide is used as the score at each step, as shown in Equation 12:

$$G(s_t, a_t) = \sum_{t=0}^{T-1} r(s_t, a_t) = G(\tau). \tag{12}$$

The $J(\theta)$ can be represented as:

$$J(\theta) = G(\tau)\left(\sum_{t=0}^{T-1} \log \pi_\theta(s_t|a_t)\right). \tag{13}$$

Combining with Equation 11, we obtain:

$$J(\theta) = G(\tau)\left(\sum_{t=0}^{T-1} \log \pi_\theta(s_t|a_t)\right) = G(\tau)p(x|c). \tag{14}$$

Olivecrona et al.(*83*) proposed a target function that can improve the efficiency of RL fine-tuning compared to the target function in Equation 14. The specific steps are as follows: (1) Copy the trained distillation model as an initialization agent model, sample peptide strings on the agent model, and save the likelihood value as $\log P(S)_{Agent}$; (2) Input the generated peptide strings into the trained distillation model to obtain the likelihood value of the distillation model $\log P(S)_{Distilled}$; (3) Multiply the reward $Score(S)$ calculated by the AMP-MIC by a coefficient $\sigma$, and add $\log P(S)_{Distilled}$ as the augmented likelihood $\log P(A)_{Agent}$.

$$\log P(A)_{Aug} = \log P(A)_{distilled} + \sigma S(A). \tag{15}$$

The loss function is defined as the augmented likelihood subtracted by the square

of the likelihood of the agent model:

$$Loss = \left(\log P\,(A)_{Aug} - \log P\,(A)_{Agent}\right)^2. \tag{16}$$

Setting $G(\tau) = \left(\log P\,(A)_{Aug} - \log P\,(A)_{Agent}\right)^2 / \log P\,(A)_{Aug}$, we can obtain Equation 17 by combining with Equation 14:

$$J(\theta) = Loss = \left(\log P\,(A)_{Aug} - \log P\,(A)_{Agent}\right)^2. \tag{17}$$

Finally, the loss function can be optimized through gradient descent algorithm. In this study, we defined the reward as:

$$G(\tau) = \omega_1 CP - \omega_2 MIC - \omega_3 SL - \omega_4 MS + \omega_5 Charge. \tag{18}$$

The variables involved in the equation have been previously explained in the section titled **Designing high-activity peptides against Gram-negative bacteria**.

**AMP-MIC**

We fine-tuned the AMP-GPT on the MIC datasets of three bacterial species, namely Staphylococcus aureus (*S. aureus*), Escherichia coli (*E. coli*), and Pseudomonas aeruginosa (*P. aeruginosa*), and compared it with GPT and LSTM prediction models.

**CT and FT models**

To validate the effectiveness of prompt learning, we used the condition token (CT) and traditional full-model fine-tuning (FT) methods as the additional baselines in this study. These models used the same backbone as AMP-GPT and employ full model fine-tuning and conditional token for training and generation. As shown in **Fig. S5**, the condition token method is retrained on labeled data and a property label is added to the front of the model during training, which is used as the input during generation to autoregressively generate the entire peptide sequence. The traditional fine-tuning directly fine-tunes the AMP-GPT model using the real AMPs dataset.

**Dataset**

The dataset employed in this study comprises of three distinct components. Firstly, the training set of AMP-GPT was assembled from UniProt(*84*), whereby the peptides of length less than 32 were selected, followed by data cleaning processes to retain only the peptides with 20 amino acids. This process led to a final set of 630,683 peptides. The second component involved the use of label data to facilitate contrastive prompt learning, which was sourced from APD3(*85*), CAMP(*86*), DBAASP(*87*), DBAMP(*16*), and DRAMP(*88*). We retained monomers and natural AMP sequences, obtaining a total of 9,896 AMPs. Subsequently, we filtered the DBAASP database for monomers and natural AMPs with target activities ≥100, resulting in 2,303 non-AMPs. The third component was the dataset employed to train AMP-MIC, which integrated MIC values. In this part, we utilized the Gramapa(*89*) dataset, which featured over 51,345 laboratory peptide results. By processing the data using the code provided by Huang et al.(*64*), we obtained 4,578 positive MIC data against *S. aureus,* 5,102 positive MIC data against *E. coli*, 2,853 positive MIC data against *P. aeruginosa* and 8048 negative data. The data was then split into training, validation, and testing sets in a ratio of 75:15:10.

**Materials**

The AMPs used in this study were custom synthesized by Nanjing TGpeptide Biotechnology Co., Ltd. and their accurate molecular weights were determined by mass spectrometry. The purity of all peptides was determined by high-performance liquid chromatography (Column: kromasil C18, 5μm, 4.6*150 mm; Wavelength: 214 nm), and all purity was greater than 95%. The abbreviations of AMPs and their corresponding sequences are shown in **Table S8**.

Cyclophosphamide, Norfloxacin and Triton X-100 were purchased from Shanghai Macklin Biochemical Co., Ltd. Colistin was obtained from Bidepharm. Isoflurane was purchased from Shandong Ante Herding Technology Co., Ltd.

All of the animal experiments performed in this research were approved by the Institutional Animal Care and Use Committee of the Guangzhou Institute of Biomedicine and Health (approval number: 2023079).

**Bacteria**

The standard bacterial strains utilized in this work including *Enterococcus faecalis*

ATCC 29212, *Staphylococcus aureus* ATCC 29213, *Klebsiella pneumoniae* ATCC 700603, *Acinetobacter baumannii* ATCC 19606, *Pseudomonas aeruginosa* ATCC 27853 and *Escherichia coli* ATCC 25922 were provided by microbiology laboratory of General Hospital of Southern Theatre Command. *Propionibacterium acnes* ATCC 6919 and ATCC 11827 were purchased from Guangdong Microbial Culture Collection Center. Pan-drug resistant (PDR) *Klebsiella pneumoniae* 418015, Extensive-drug resistant (XDR) *Klebsiella pneumoniae* 325016, Multi-drug resistant (MDR) *Klebsiella pneumoniae* 327004, MDR *Escherichia coli* 103231, MDR *Pseudomonas aeruginosa* 304238 and MDR *Acinetobacter baumannii* 316039 were isolated clinically from patients in the General Hospital of Southern Theatre Command and their antimicrobial susceptibility testing results are displayed in **Table S6**. All strains of ESKAPE pathogens were cultured aerobically at 37°C in cation-adjusted Mueller-Hinton broth (CAMHB) or Mueller-Hinton agar (MHA), while *Propionibacterium acnes* were cultured anaerobically in Brain-Heart Infusion (BHI) broth or anaerobic blood agar.

**Mice**

Male BALB/c mice (6-8 weeks, 18-22 g) were purchased from Guangdong Medical Laboratory Animal Center. All animal experiments were performed according to the 'Principles of Laboratory Animal Care' (NIH publication No. 86–23, revised 1985) and approved by the Institutional Animal Care and Use Committee.

**MIC determination**

The MICs of AMPs against Gram-negative bacteria and aerobic Gram-positive bacteria were determined by broth microdilution method according to the guidelines of the Clinical and Laboratory Standards Institute (CLSI). Briefly, AMPs were dissolved into sterile distilled water with an initial concentration of 5120 μg/mL and stored at 4°C. In a sterile, plastic 96-well U-bottom plate, using a multichannel pipette, 100 μL of CAMHB were added in wells of columns 2–12, subsequently, 200 μL of the peptide solutions diluted to 1:10 with a final concentration of 512 μg/mL were added to the column 1, and a series of twofold dilutions were prepared by transferring 100 μL to successive wells. The final test concentration of AMP was 256-0.125 μg/ml, three replicates of each peptide concentration were tested against each bacterial strain. A single colony of the bacterial strain was transferred to 4 mL of CAMHB and cultured at 37 °C. After about 4 h, the bacterial culture at the logarithmic phase was adjusted to 0.5 McFarland and diluted 1:100 with fresh CAMHB, and 100 μL was added to 96-well plates containing AMPs to a final density of $5 \times 10^5$ CFU/mL. Each test included a sterile blank control (CAMHB only), a growth (negative) control (CAMHB with bacterial inoculums but no peptide/antibiotic) and a positive control (colistin against *E. coli* and *P. aeruginosa*, vancomycin against *E. faecalis* and *S. aureus*, meropenem against *K. pneumoniae* and *A. baumannii*). After incubating at 37°C for 16-20 h, the MIC results were taken as the minimal concentration of AMPs at which no visible bacterial growth was observed.

The MICs of AMPs against anaerobic Gram-positive bacteria were determined as described in He et al(*90*). The *P. acnes* were inoculated into BHI medium and incubated in an anaerobic bag at 37°C to reach the logarithmic growth phase. The bacterial suspensions were prepared through centrifuged and resuspended in BHI medium three times, and then were diluted to a final concentration of $1 \times 10^6$ CFU/ml. In a sterile, plastic 96-well U-bottom plate, using a multichannel pipette, 100 μL of CAMHB were added in wells of columns 2–11, subsequently, 200 μL of the peptide solutions diluted to 1:20 with a final concentration of 128 μg/mL were added to the column 1, and a series of twofold dilutions were prepared by transferring 100 μL to successive wells. The final test concentration of AMP was 64-0.0625 μg/ml, three replicates of each peptide concentration were tested against each bacterial strain. Then, 100 μL aliquots of bacterial suspensions were added to each well. Each test included a sterile blank control (BHI only), a growth (negative) control (BHI with bacterial inoculums but no peptide/antibiotic) and a positive control (clindamycin). The MIC values were obtained as the lowest drug concentrations observed without any evidence of bacterial growth after 48 h anaerobic culture.

**Time-kill assay**

The *E. coli* culture, adjusted to 0.5 McFarland, was diluted 1:100 with fresh CAMHB medium and incubated at 37°C, 150 r.p.m. for 3-4 h to reach the logarithmic growth phase. The bacterial solution was then adjusted to 1 McFarland and diluted 1:100 in 10 mL CAMHB medium containing the appropriate concentration of AMPs (16 μg/mL of KW13, 32 μg/mL of AI18 and 2 μg/mL of COL) and incubated at 37°C, 150 r.p.m. The time at which the bacterial solution was added was defined as 0 h. At 0, 2, 4, 6, 18 and 24 h, 200 μL of bacteria solutions were taken out and were ten-fold serially diluted with fresh CAMHB medium in a sterile, 96-well U-bottom plate. Next, 10 μL of each dilution was dropped onto MHA plates and incubated at 37°C for 18-24 h to read the results. The assays were performed with three independent replicates. The lower limit of detection for the time-kill studies was 300 CFU/ml. These are based on dropping 10ul of bacterial solution and counting only dilutions of ≥3 and ≤30 CFUs.

**Hemolysis test**

Fresh human red blood cells (RBCs) were washed three times with sterile PBS (800 g, centrifuged at 4°C for 5 min), and then added to 96-well U-bottom plates containing AMPs, making a concentration of 2% for RBCs and 100 μg/mL for AMPs, with a final volume of 200 μL. PBS treated RBCs were used as negative control and 0.5% Triton X-100 treated RBCs were used as positive control. After incubation at 37°C for 1 h, the mixtures were centrifuged at 1200 g for 15 min at 4°C. The supernatants were collected in new 96-well plates with a flat bottom and the $OD_{570}$ was measured. All experiments were performed with three independent replicates. The rate of hemolysis is calculated according to the following equation:

$$hemolysis\ (\%) = \frac{OD_{570,sample} - OD_{570,negative}}{OD_{570,positive} - OD_{570,negative}}. \qquad (19)$$

**Resistance development in *E. coli***

To evaluate the resistance development during serial passage, *E. coli* ATCC 25922 cells were cultured to logarithmic growth phase, then adjusted to 0.5 McFarland and diluted 1:100 into fresh CAMHB, then 100 μL of bacteria suspentions were added to 96-well plates containing two-fold serial dilutions of varying concentrations of AMPs to a final volume of 200 μL, MIC was obtained after incubating at 37 °C without shaking for 24 h. The cells that can grow in the wells with the highest concentration of AMPs were diluted 1:10000 into new 96-well plates containing fresh CAMHB and two-fold serial diluted AMPs with varying concentrations again. The same procedure was performed every 24 h and stopped after 30 days. Norfloxacin and COL were selected as control antibiotics. The change in MIC value was recorded by taking the normalized value of MIC at generation X against the MIC value at generation 1.

**Plasma stability test**

The sterile human blood was centrifuged at 3500 rpm for 10 min to obtain the supernatant as freshly prepared plasma. AMP was dissolved in water and freshly prepared plasma (1:1) to prepare a stock solution of 2500 μg/mL.

Triplicate tubes were gently shaken in an orbital shaker at 37 °C (100 r.p.m.) and 200 μl aliquots were collected at 0, 60, 120 and 240 min and added to 800 μl of a mixture containing 10% methanol:10% water:80% acetonitrile and 1% acetic acid to stop further degradation of the peptides. The suspensions were centrifuged and the supernatants were collected, filtered using a 0.45 μm syringe filter, and analyzed by HPLC using the Agilent ZORBAX C18 reverse-phase analytical column (5 μm, 4.6 × 250 mm). The mobile phases were acetonitrile (containing 0.1% TFA) and deionized water (containing 0.1% TFA), and the elution system used was from 10 to 99% acetonitrile, the analytic time was 33 min, the flow rate was 1 mL/min, and absorbance was measured at a wavelength of 214 nm.

**Membrane disruption detected by transmission electron microscopy**

A single colony of *E. coli* ATCC 25922 was cultured in 60 mL of CAMHB medium overnight to the logarithmic growth phase at 37 °C, then bacteria cells were collected by centrifugation at 4000 rpm for 5 min at room temperature and resuspended with PBS (pH 7.0) to an $OD_{600}$ of 1.0. 10 mL of the bacterial suspensions were treated with AMP at final concentrations of 1 × MIC and 10 × MIC at 37 °C for 5 h, untreated cells were used as control.

Samples were washed with PBS three times and prefixed with 1 mL of 3% glutaraldehyde (v/v) in PBS at 4 °C for 3 h. After washing with PBS two times in 10 min, samples were postfixed with 1% osmium acid for 2 h at room temperature and then washed with PBS two times in 5 min. Thereafter, the specimens were subsequently dehydrated in ethanol series (10 min each: 50%, 70%) and acetone series (10 min × 2: 80%, 90%, 100%). Subsequently, the specimens were displaced in a mixture of acetone and epoxy resin Epon812 (1:1 for 40 min) and infiltrated in epoxy resin Epon812 overnight at 37 °C and finally embedded in epoxy resin Epon812 at 60°C for 48h in a cylindrical mold. Ultrathin sections of 40 nm thickness using Leica ultramicrotome were then carefully transferred to a copper grid and dried overnight. These copper grids were post-stained with saturated uranyl acetate (prepared with 70% ethanol) for 3 min and lead citrate for 3min and dried completely. The images were taken with a TEM (HITACHI H-7650).

**Neutropenic mouse pneumonia model**

Twenty-four healthy male BALB/c mice (6-8 weeks, 20±2 g) were injected intraperitoneally with cyclophosphamide (dissolved in sterile PBS) 4 d and 1 d before bacterial infection, at concentrations of 150 mg/kg and 100 mg/kg, respectively, to induce neutropenia. On the day of infection, mice were anaesthetized with isoflurane and then suspended by their upper front teeth from a nylon string on an angled plexiglass platform. 40 μL of log-phase grown *K. pneumoniae* 325016 ($1 \times 10^6$ CFU/ml in PBS), loaded in a pipette were instilled into the lungs by compressing air. Mice were hold upright for 30 s to allow bacteria inoculum to be inhaled into the lungs. Subsequently, the mice were randomly divided into 4 groups of 6 mice each. The mice were injected intraperitoneally with 10 mg/kg of KW13, AI18, indolicidin or an equivalent amount of PBS at 2 h and 12 h post-infection, respectively. All mice were euthanized at 24 h after infection, and the left lung tissues were collected under aseptic conditions, weighed and homogenized with 1 ml of sterile PBS. The homogenate was diluted with sterile PBS in a ten-fold gradient and plated onto MHA plates, incubated at 37°C for 18-24h, then colonies were counted and the bacterial load of lung tissue was calculated.

Histological analyses were conducted after the right lung tissues were collected and fixed by 4% formalin overnight and embedded in paraffin for hematoxylin and eosin (H&E) staining.

**Statistical Analysis**

Statistical analysis was conducted using one-way analysis of variance. The data were analyzed using statistical software of GraphPad Prism version 8.0 (Chicago, IL, USA). Quantitative data were expressed as mean ± standard deviation (SD), and $P < 0.05$ was considered statistically significant.

## Acknowledgements

**Funding:** This work was financially supported by National Natural Science Foundation of China (92370130, 82204279, 81973281, 82373791, 22303083), National Key Research and Development Program of China (2022YFF1203003), Fundamental Research Funds for the Central Universities (226-2022-00220), China Postdoctoral Science Foundation (2023M733128, 2023TQ0285), Postdoctoral Fellowship Program

**Author contributions:**

Project administration: C.Y.H., Z.H.J., and T.J.H.

Conceptualization: J.K.W., Y.K., X.C.Z., C.Y.H., Z.H.J. and T.J.H.

Data curation: J.K.W., M.Y.W., G.G.Y., and Z.H.J.

Formal analysis: J.K.W., J.W.F., G.G.Y., M.Y.W., Z.X.W., Y.J.M., L.R.W., Y.W., J.X.G., T.Y.W., and Z.H.J.

Investigation: J.K.W., J.W.F., Y.J.M., L.R.W., Y.W., Z.H.J., and T.J.H.

Methodology: J.K.W., J.W.F., Y.K., M.Y.W., Z.H.J., and T.J.H.

Resources: J.K.W., J.W.F., Y.K., P.C.P., Y.J.M., L.R.W., Y.W., C.Y.H., Z.H.J., and T.J.H.

Funding acquisition: J.K.W., Y.K., P.C.P., Z.H.J., and T.J.H.

Visualization: J.K.W., J.W.F., Y.K., X.C.Z., Z.H.J.,

Software: J.K.W.

Supervision: Y.K., P.C.P., H.S., Y.F.D, C.Y.H., Z.H.J., and T.J.H.

Validation: J.K.W., J.W.F., L.R.W., Y.W., X.C.Z., and Z.H.J.,

Writing—original draft: J.K.W., J.W.F., Y.K., M.Y.W., Z.H.J., and T.J.H.

Writing—review and editing: J.K.W., J.W.F., Y.K., P.C.P., M.Y.W., X.J.Z., X.C.Z., C.Y.H., Z.H.J., and T.J.H.

**Competing interests:** The authors declare that they have no competing interests.

**Data and materials availability:** All data needed to evaluate the conclusions in the paper are present in the paper and/or the Supplementary Materials. The dataset utilized in this investigation consists of three components. Firstly, the UniProt(*84)* database was employed to retrieve peptides with a length less than 32. The second component involved label data obtained from multiple sources, namely APD3(*85*), CAMP(*86*),

DBAASP(*87*), DBAMP(*16*), and DRAMP(*88*). The third component encompasses the MIC data, sourced from the Gramapa(*89*) dataset. The code used in the study can be found at https://github.com/jkwang93/AMP-Designer or at https://doi.org/10.5281/zenodo.13999503.

## Supplementary Materials

**Diversity and novelty analysis between Prompt, FT and CT**

In order to conduct a comprehensive comparison among the prompt tuning, fine-tuning, and condition token models, we presented the match scores using statistical tables and cumulative distribution curves, as depicted in **Table S2** and **Fig. S1a-b**, respectively. **Table 2** provides a comprehensive statistical analysis of the mean match scores across different intervals. Notably, in the comparison of match scores with real AMPs, the average match scores for the Prompt, FT, and CT models were determined at 6.53, 15.58 and 13.12, respectively. Similarly, in the comparison of Internal Match Scores, the average match scores for the Prompt, FT, and CT models were observed to be 5.64, 13.80 and 11.63, respectively. These results offer a clearer understanding of the models' performance and highlight the varying degrees of similarity between the generated sequences and both Real-AMPs and internal sequences. Moreover, when examining **Fig. S1a-b**, it becomes apparent that across almost all intervals, the Prompt model consistently outperforms the FT and CT models in terms of match scores.

To further validate the diversity of AMPs generated by the three models, we computed the distribution of entropy, which reflects the uncertainty of the model's predictions at each time step. Higher entropy values indicate greater uncertainty and a larger number of possible outputs, while lower entropy values indicate higher certainty and fewer possible outputs for the model's predictions. As illustrated in **Fig. S2**, it is evident that the prompt model exhibits higher entropy compared to the FT and CT models. This indicates that the FT and CT models generate sequences with higher certainty compared to the prompt model, resulting in lower diversity of generated sequences.

The aforementioned findings provide compelling evidence of the significant distribution adjustment achieved through prompt tuning the original model. This adjustment ensures that the model, following prompt tuning, is capable of generating candidate sequences with desired activity. Moreover, compared to the fine-tuning and condition token models, the prompt tuning approach generates AMPs with lower similarity to real AMPs, thereby exhibiting higher diversity. These results underscore the efficacy of prompt tuning in shaping the generated sequences, enhancing their dissimilarity to real AMPs, and promoting increased diversity in the generated AMPs.

**Time and memory space usage**

In this section, we present a comparative analysis of the memory space and time utilization of AMP-GPT AMP-prompt and AMP-Distillation. As shown in **Table S3**, the "Model Size" column represents the memory space usage occupied by each model, while the "Time" column indicates the time required to generate 5000 AMPs. The slower generation speed of the AMP-prompt model can be attributed to the concatenation of prompt embeddings at the beginning. In comparison to the other two models, the AMP-Distillation model has significantly smaller model size and nearly a tenfold improvement in speed compared to the AMP-Prompt model. The tests were conducted on a computer equipped with an AMD Ryzen 7 5800H CPU, an NVIDIA GeForce RTX 3060 Laptop GPU, and 16 GB of RAM.

**Training information**

The model is constructed using the huggingface's transformers package, and the hyperparameters of the model are as follows: "activation_function": "gelu_new", "architectures": ["GPT2LMHeadModel"], "attn_pdrop": 0.1, "n_ctx": 50, "n_embd": 768, "n_head": 8, "n_layer": 12. AdamW optimizer was used in training, for further details regarding additional parameters, please refer to our GitHub repository.

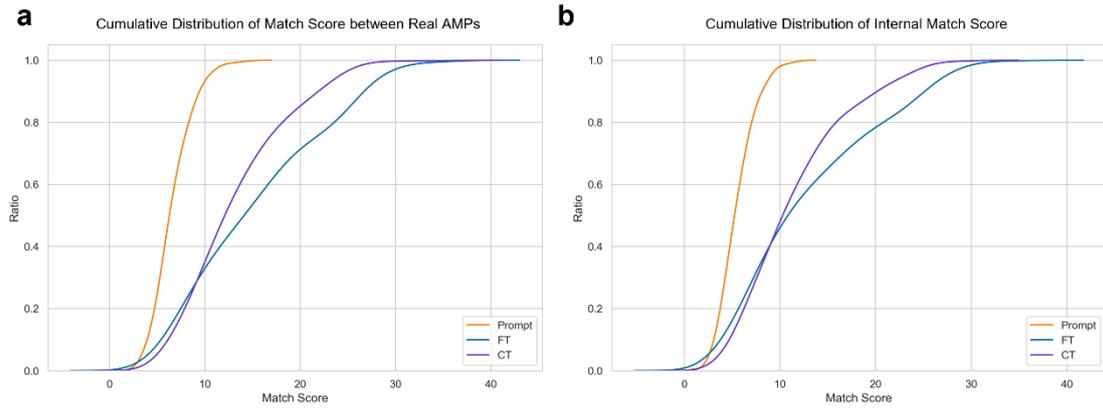

**Fig. S1. Match scores cumulative distribution curves.** Cumulative distribution curves of match score. **(a).** real AMPs and **(b).** them self.

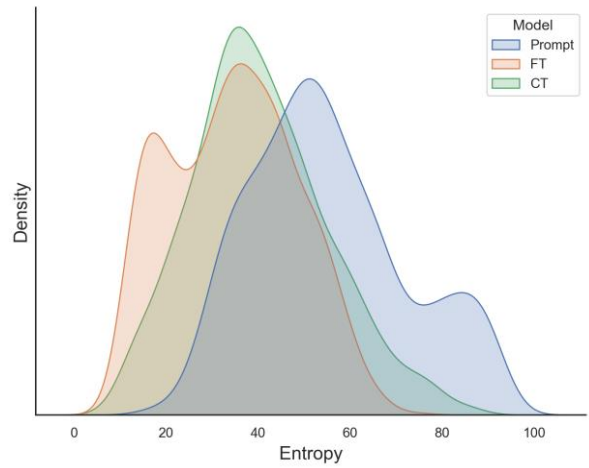

**Fig. S2. The entropy distribution of Prompt, FT and CT**.

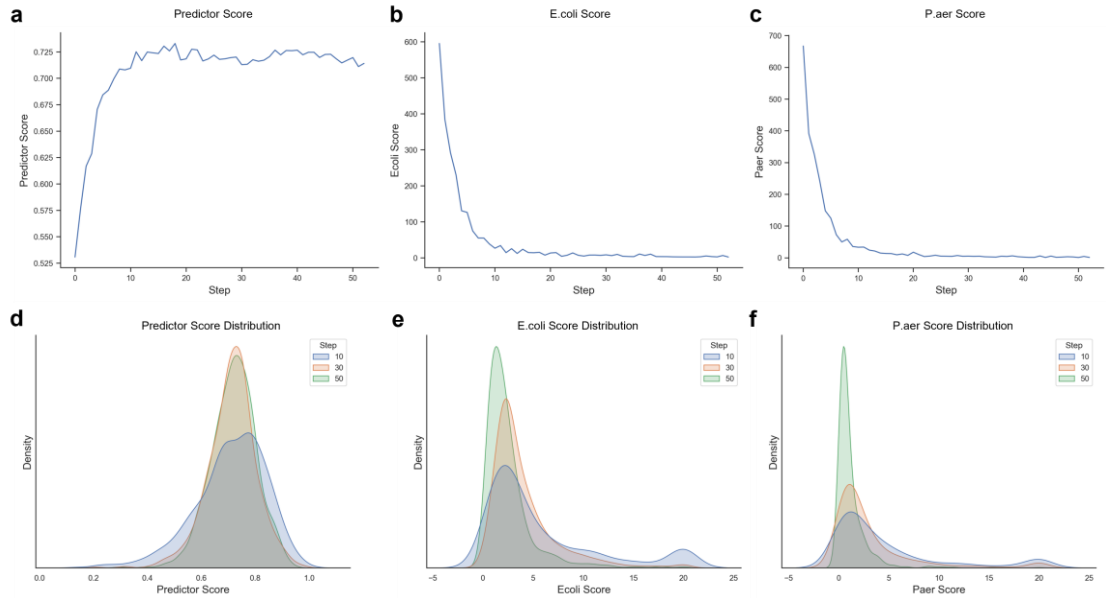

**Fig. S3. The variations of different attributes during the RL. (a-c).** The relationship between average score and step: **(a)** AMP predictor score, **(b)** *E. coli* predicted score and **(c)** *P. aeruginosa* predicted score. **(d)** AMP predictor score distribution. **(e)** *E. coli* predicted score distribution, and **(f)** *P. aeruginosa* predicted score distribution.

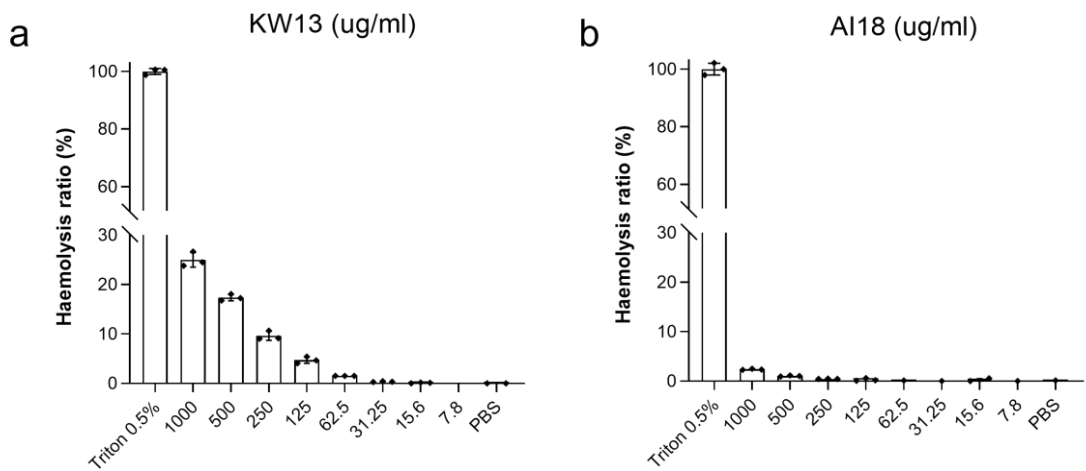

**Fig. S4. Hemolytic concentration.** 50% hemolytic concentration (HC50, 50% hemolysis) of KW13 and AI18.

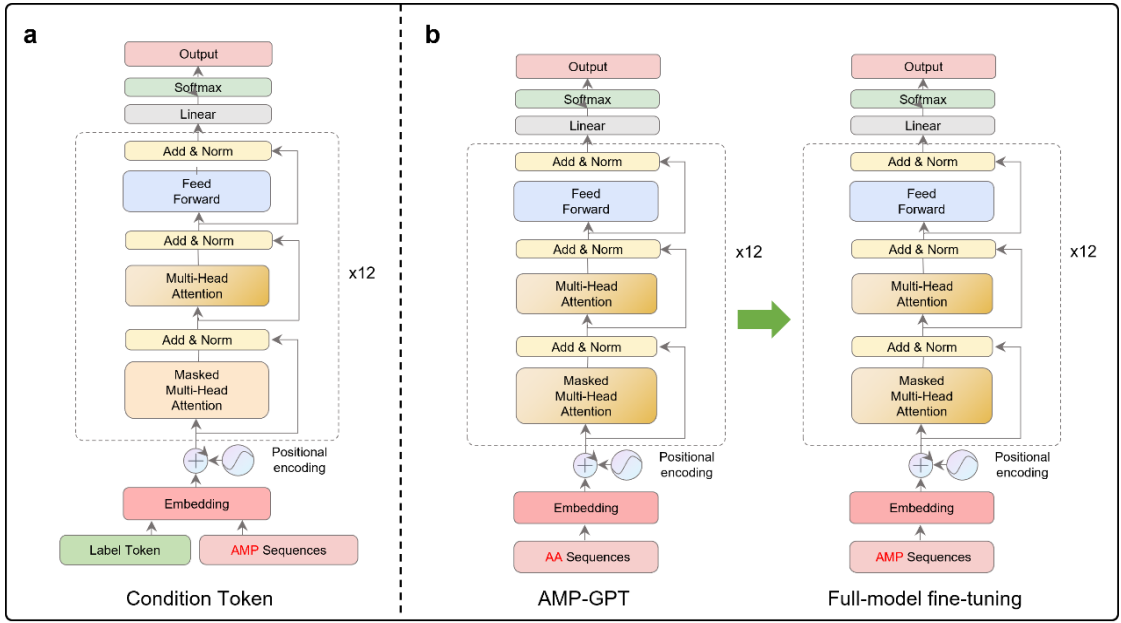

**Fig. S5. The model architecture. (a).** CT and **(b).** FT.

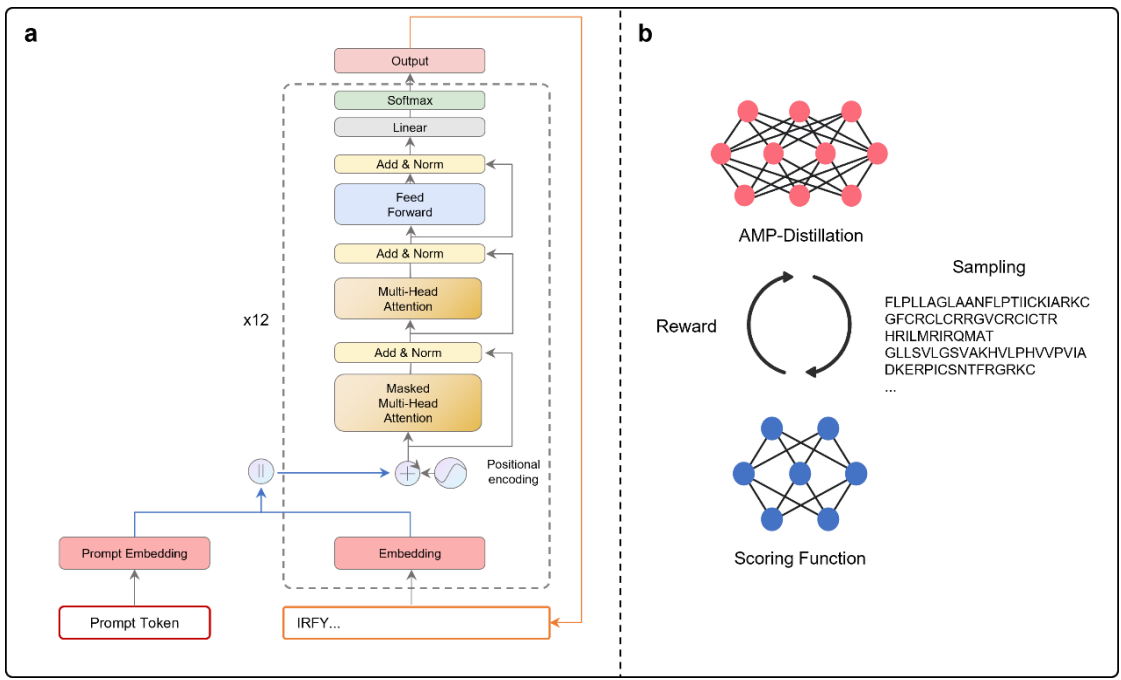

**Fig. S6. The Inference stage. (a).** AMP-Prompt and **(b).** RL.

**Table S1.** Percentage of the peptides identified as AMPs by all three predictors simultaneously.

| Type | Percentage |
|---|---|
| UniProt | 4.58% |
| GPT | 5.13% |
| Prompt | 61.8% |
| Prompt-TopK | 83.4% |
| Distillation | 59.2% |
| Real-AMPs | 81.4% |

**Table S2.** Match scores distribution.

| | Model | Mean | Std | 25% | 50% | 75% |
|---|---|---|---|---|---|---|
| Self | AMP-GPT | 1.80 | 2.63 | 0.81 | 1.20 | 1.99 |
| | AMP-Prompt | 5.64 | 1.93 | 4.21 | 5.50 | 6.83 |
| | AMP-Prompt-TopK | 7.29 | 1.93 | 6.00 | 7.09 | 8.40 |
| | AMP-Distillation | 6.09 | 1.95 | 4.73 | 6.11 | 7.30 |
| | Condition Token (CT) | 11.63 | 5.68 | 7.21 | 10.47 | 14.75 |
| | Fine-Tuning (FT) | 13.80 | 7.96 | 7.09 | 11.71 | 19.67 |
| Real AMP | AMP-GPT | 4.17 | 3.19 | 2.79 | 3.53 | 4.43 |
| | AMP-Prompt | 6.53 | 2.13 | 5.08 | 6.23 | 7.87 |
| | AMP-Prompt-TopK | 7.43 | 2.15 | 5.89 | 7.11 | 8.62 |
| | AMP-Distillation | 6.89 | 2.13 | 5.38 | 6.60 | 8.11 |
| | Condition Token (CT) | 13.12 | 5.72 | 8.65 | 12.14 | 16.61 |
| | Fine-Tuning (FT) | 15.58 | 7.90 | 8.40 | 14.59 | 23.05 |

**Table S3.** Time and space comparison of AMP-Distillation and AMP-prompt and AMP-GPT using.

| Methods | Model Size(MB) | Time(S) |
|---|---|---|
| AMP-GPT | 334 | 51.9 |
| AMP-prompt | 335 | 136.8 |
| AMP-Distillation | 16 | 11.9 |

**Table S4.** Reward of RL.

| Reward | Detail |
|---|---|
| Classification probably ($CP$) | Classification probably predicted by Macrel and divided into 5 bins. |
| MIC of three bacteria ($MIC$) | MIC predicted by AMP-MIC divided into 10 bins. |
| Sequence Length ($SL$) | Sequence length of peptides. |
| Memory Similarity ($MS$) | Similarity between generated sequences and AMPs in memory queue. |
| Charge | Continuous value. |

**Table S5.** The performance of the MIC predictors.

| Bacteria | Metric | RNN | GPT | AMP-MIC |
|---|---|---|---|---|
| S. aureus | MSE | 0.424 | 0.394 | **0.373** |
|  | $R^2$ | 0.792 | 0.801 | **0.811** |
| E. coli | MSE | 0.464 | 0.413 | 0.393 |
|  | $R^2$ | 0.771 | 0.796 | **0.807** |
| P. aeruginosa | MSE | 0.323 | 0.328 | **0.263** |
|  | $R^2$ | 0.816 | 0.813 | **0.851** |

**Table S6.** Antimicrobial susceptibility testing results of clinical bacteria.

| Antibiotic | MIC (μg/mL) and susceptibility profile | | | | | |
|---|---|---|---|---|---|---|
| | 316039 | 304238 | 103231 | 327004 | 325016 | 418015 |
| Ticarcillin/clavulanic acid | 128 R | 128 R | 32 R | 128 R | 128 R | 128 R |
| Piperacillin/tazobactam | 128 R | ND | 128 R | 128 R | 128 R | 128 R |
| Ceftazidime | 64 R | 32 R | 32 R | 64 R | 64 R | 64 R |
| Ceftriaxone | ND | ND | 64 R | ND | ND | ND |
| Cefoperazone/sulbactam | 64 R | 64 R | 64 R | 64 R | 64 R | 64 R |
| Cefepime | 32 R | 16 I | 32 R | 32 R | 32 R | 32 R |
| Aztreonam | ND | ND | ND | 64 R | 64 R | 64 R |
| Imipenem | 16 R | 8 R | 0.25 S | 16 R | 16 R | 16 R |
| Meropenem | 16 R | 4 I | ND | 16 R | 16 R | 16 R |
| Amikacin | ND | 2 S | 16 S | 64 R | 64 R | 64 R |
| Tobramycin | 1 S | 1 S | ND | 16 R | 16 R | 16 R |
| Ciprofloxacin | 4 R | 1 I | ND | 4 R | 4 R | 4 R |
| Levofloxacin | 4 I | 4 R | 8 R | 8 R | 8 R | 8 R |
| Doxycycline | 16 R | ND | ND | 2 S | 16 R | 8 I |
| Minocycline | 4 S | ND | ND | 4 S | 16 R | 16 R |
| Colistin | 0.5 S | 0.5 S | 0.5 S | 0.5 S | 0.5 S | 16 R |
| Sulfamethoxazole/trimethoprim | 160 R | 320 R | 320 R | 320 R | 20 S | 320 R |
| Tigecycline | 2 S | - | 0.5 S | 0.5 S | 8 R | 2 R |

R, resistant; S, sensitive; I, intermediate. -, intrinsic resistance. ND, not determined.

**Table S7.** The comparison with AMPs in analogous work.

| Method | ID | MIC[a] (μg/ml) | | | Coverage% | Similarity% | E-value | BLAST score |
|---|---|---|---|---|---|---|---|---|
| | | *S. aureus* | *E. coli* | *P. aeruginosa* | | | | |
| Our | RR13 | 64 | 16 | 32 | 53.8 | 100 | 0.35 | 30.33 |
| | AI18 | 32 | 8 | 8 | 77.8 | 71.4 | 1.07 | 31.60 |
| | KW20 | 16 | 4 | 8 | 65.0 | 76.9 | 0.61 | 30.75 |
| | KW13 | 8 | 4 | 4 | 84.6 | 75.0 | 0.30 | 30.33 |
| | VK18 | 16 | 16 | 16 | 55.6 | 80.0 | 0.15 | 32.03 |
| | SK18 | 8 | 8 | 16 | 83.3 | 80.0 | 0.47 | 30.75 |
| | RV15 | 8 | 8 | 16 | 73.3 | 72.7 | 1.05 | 29.91 |
| | IR20 | 32 | 32 | 64 | 80.0 | 63.2 | 0.07 | 33.72 |
| | RK17 | 32 | 32 | 16 | 64.7 | 81.8 | 0.80 | 32.03 |
| | LW14 | 16 | 64 | 16 | 78.6 | 69.2 | 0.01 | 34.57 |
| Das et al. [30][b] | YI12 | 8 | 31 | 125 | 66.7 | 87.5 | 1.20 | 29.90 |
| | FK13 | 16 | 31 | 63 | 69.2 | 88.9 | 0.03 | 32.88 |
| Huang et al.[63][c] | CRRI3 | 8 | 64 | 16 | 100 | 100 | 0.87 | 30.33 |
| | CRRI4 | 8 | 64 | 64 | 100 | 100 | 0.12 | 30.33 |
| | CRRI7 | 8 | 32 | 32 | 100 | 100 | 0.03 | 32.03 |

[a] Minimal inhibitory concentration; [b] *Staphylococcus aureus* (ATCC 29737), *Escherichia coli* (ATCC 25922), *Pseudomonas aeruginosa* (ATCC 9027); [c] *Staphylococcus aureus* (ATCC 6538), *Escherichia coli* (ATCC 1806), *Pseudomonas aeruginosa* (ATCC 2512).

**Table S8.** The abbreviations of AMPs and their corresponding sequences.

| Abbreviation | Sequence | Expected MS | Measured MS |
|---|---|---|---|
| RR13 | RRWFKIRMAAKLA-NH2 | 1646.08 | 1645.88 |
| AI18 | AIPKRLRRFYLRALARRL-NH2 | 2268.84 | 2269.44 |
| KW20 | KWRKVRAKFWRKWLAGLINT-NH2 | 2557.14 | 2557.84 |
| KW13 | KWKKWVRAIKKLV-NH2 | 1682.18 | 1682.40 |
| VK18 | VKRFRRWWKPWRKILHLV-NH2 | 2505.13 | 2505.32 |
| SK18 | SKLWKKIVKALKAALKTL-NH2 | 2038.64 | 2039.00 |
| RV15 | RVLWIKRWIKRFFRP-NH2 | 2100.64 | 2101.00 |
| IR20 | IRFYWRRVQVWRGIWRRLVR-NH2 | 2801.40 | 2801.20 |
| RK17 | RKKWWAYLLAKIAKKVK-NH2 | 2129.72 | 2129.94 |
| LW14 | LWRFKRWWWWKKIL-NH2 | 2132.65 | 2132.80 |
| GI16 | GIKKFAKLLKYIAGKL-NH2 | 1790.29 | 1790.55 |
| RR16 | RRWGKLIKKIAKKFGG-NH2 | 1885.35 | 1885.00 |
| YK19 | YKWVANVAKKLIKLLKVLK-NH2 | 2254.89 | 2254.80 |
| AK15 | AKKFKIWGAIKRLLA-NH2 | 1742.20 | 1741.95 |
| KK12 | KKFLKLIKGIKR-NH2 | 1470.93 | 1471.05 |
| KWI18 | KWKWPPRWPPWRPVWKVI-NH2 | 2441.96 | 2441.85 |
| LF19 | LFKVIFGIVKKKKLLPKFF-NH2 | 2292.98 | 2293.00 |
| AV20 | AVWRWLWKGSLAKGIIKFLK-NH2 | 2399.96 | 2400.40 |
| FI19 | FIKKLLKLLGKLARLAHAL-NH2 | 2145.80 | 2146.16 |
| IF20 | IFKLLKKIIKKIVFKIAKAL-NH2 | 2355.18 | 2355.16 |
| RW20 | RWPFKWKRPRILKLICFRAL-NH2 | 2628.34 | 2628.80 |
| RW12 | RWKVIRRRWLRL-NH2 | 1738.18 | 1738.20 |
| KW18 | KWKWWKRRKFCLLYIVFC-NH2 | 2504.16 | 2504.40 |